\documentclass[a4paper,10pt,]{article}

\usepackage{amsmath}
\usepackage{amssymb}
\usepackage{graphicx}
\usepackage{latexsym}
\usepackage{color}
\usepackage[colorlinks=true,linkcolor=red]{hyperref}%

\usepackage{authblk}

\usepackage{mathabx}

\newtheorem{thm}{Theorem}[section]
\newtheorem{defi}[thm]{Definition}
\newtheorem{lemma}[thm]{Lemma}
\newtheorem{prop}[thm]{Proposition}
\newtheorem{corol}[thm]{Corollary}

\newtheorem{remark}[thm]{Remark}
\newtheorem{example}[thm]{Example}

\newcommand{\eps}{\varepsilon}
\newcommand{\fie}{\varphi}
\newcommand{\sipa}[1]{\big(\,#1\,\big)}
\newcommand{\cp}{\mathbb{P}^1(\mathbb{C})}
\newcommand{\cpp}{\mathbb{P}^1\times\mathbb{P}^1}
\newcommand{\pik}[1]{{\rm Pic}\,#1}

\newcommand{\pikq}[1]{{\rm Pic}_\mathbb{Q}\, #1}
\newcommand{\PP}[1]{{\rm P}_{\!#1}}
\newcommand{\pfie}{\overline\fie_*}
\newcommand{\ve}[2]{\begin{pmatrix} #1 \\  #2 \end{pmatrix}}

\makeatletter
\newcommand{\xleftrightarrow}[2][]{\ext@arrow 3359\leftrightarrowfill@{#1}{#2}}
\newcommand{\xdashrightarrow}[2][]{\ext@arrow 0359\rightarrowfill@@{#1}{#2}}
\newcommand{\xdashleftarrow}[2][]{\ext@arrow 3095\leftarrowfill@@{#1}{#2}}
\newcommand{\xdashleftrightarrow}[2][]{\ext@arrow 3359\leftrightarrowfill@@{#1}{#2}}
\def\rightarrowfill@@{\arrowfill@@\relax\relbar\rightarrow}
\def\leftarrowfill@@{\arrowfill@@\leftarrow\relbar\relax}
\def\leftrightarrowfill@@{\arrowfill@@\leftarrow\relbar\rightarrow}
\def\arrowfill@@#1#2#3#4{%
  $\m@th\thickmuskip0mu\medmuskip\thickmuskip\thinmuskip\thickmuskip
   \relax#4#1
   \xleaders\hbox{$#4#2$}\hfill
   #3$%
}
\makeatother

\newcommand\ddaaux[1]{\rotatebox[origin=c]{-90}{\scalebox{#1}{$\dashrightarrow$}}} 
\newcommand\dashdownarrow[1]{\mathrel{\text{\ddaaux{#1}}}}

\newcommand\changes[1]{\textcolor{black}{#1}}

\begin{document}

\title{Singularity confinement as an integrability criterion}
\author[1]{\large Takafumi Mase}
\author[1]{Ralph Willox\thanks{willox@ms.u-tokyo.ac.jp}}
\author[2]{Alfred Ramani}
\author[2]{Basil Grammaticos}
\affil[1]{Graduate School of Mathematical Sciences, the University of Tokyo, 3-8-1 Komaba, Meguro-ku, 153-8914 Tokyo, Japan}
\affil[2]{ IMNC, Universit\'{e} Paris VII \& XI, CNRS, UMR 8165, Orsay, France}

\maketitle

\begin{abstract}
In this paper we present a rigorous method for deciding whether a birational three point mapping that has the singularity confinement property is integrable or not, based only on the structure of its (confined) singularity patterns. We also explain how the exact value of the dynamical degree for such a mapping may be deduced from the singularity patterns.
\end{abstract}

\section{Introduction}\label{intro}
Over the last decade, the dynamical degree (or the algebraic entropy, its logarithm) has become the principal integrability criterion in the study of (bi-)rational mappings. If a mapping (be it autonomous or not) has a dynamical degree greater than 1 it is deemed nonintegrable, whereas if its dynamical degree is exactly 1 it is said to be integrable. Because of its deep connections, not only to the original notion of complexity as introduced by Arnold \cite{arnold,veselov}, but also to the singularity structure and the underlying geometric properties of such mappings and also, ultimately, to the existence or absence of nontrivial symmetries for them \cite{cantat-ann}, the integrability criterion cited above has by now become the very definition of integrability for (bi-)rational mappings, at least in the second order case.

Although the dynamical degree of a second order birational mapping  can be calculated exactly for mappings that are {algebraically stable} \cite{dillerfavre} (possibly after blowing-up, if the mapping does not have this property originally), this can be an arduous task. 

This is why R. Halburd's method \cite{rodzero} for calculating the exact degree of the iterates of a birational (three point)  mapping that enjoys the singularity confinement property, based on its singularity structure, should be considered a milestone in the research on discrete integrable systems.
However, to decide whether the dynamical degree of a given mapping is greater than 1 or not, i.e. whether the mapping is integrable or not, it is not necessary to know the exact degree for every iterate of the mapping: It is sufficient to know how the degree of the iterates grows asymptotically. This is why we introduced a pared down version of Halburd's method \cite{rodone}, which we christened the `express' method and which is designed to yield a straightforward answer to the question whether a confining mapping is integrable or not (though not the exact degree of each iterate).
The method is extremely simple for mappings that have only a single singularity pattern. For example, the three point mapping given by the equation ($x_n\in\cp$ for all $n$)
\begin{equation}
x_{n+1} + x_{n-1} = \dfrac{1}{x_n^k},
\label{mapk}
\end{equation}
where $k\in\mathbb{Z}_{>0}$, has the singularity confinement property if and only if $k$ is even. Clearly, the only singularity of this mapping, i.e. the only value of $x_n$ where all information on $x_{n-1}$ is lost when calculating $x_{n+1}$, arises at $x_n=0$. When $k$ is an even (positive) integer, this singularity is {\it confined}. To verify this last statement, one can use the usual \cite{sincon, review} continuity argument: start from some generic initial value $x_{n-1}=u\in\cp$ and $x_n=\eps\in\mathbb{C}$, $\eps\neq0$ but $\vert\eps\vert$ small. Iterating the mapping, one then finds that
\begin{gather}
x_{n+1}\sim \eps^{-k} ,\quad x_{n+2} = -\eps + {\cal O}(\eps^{k^2}) ,\quad x_{n+3} = u + o(\eps^0),
\end{gather}
and for all subsequent iterates, $x_{n+k} = f_k(u) + o(\eps^0)$, where $f_k(u)$ $(k\geq4)$ are functions of $u$ only.
This means, by continuity of the iterates when $\eps\to0$, that after entering the singularity by $x_n=0$, one passes through values $x_{n+1}=\infty$ and $x_{n+2}=0$, after which the next iterate will again depend on the initial condition $u$. We call such a singularity {\it confined}, and associate with it the {\it confined singularity pattern}
\begin{equation}
\sipa{0~~ \infty^k~~ 0},
\label{sipak}
\end{equation}
denoting, from left to right, the entry value, the values traversed upon iteration and finally the exit value after which one recovers the information on the initial condition that was lost when entering the singularity. The exponent in the pattern is of course that obtained for the leading order term $(\eps^{-1})^{k}$ in $x_{n+1}$ in the above analysis. The reason for recording this much detail in the singularity pattern is the following. Let us associate a monomial $c_j \lambda^{j-1}$ with each entry in the pattern, where $j$ is the position of the entry (counted from left to right, starting at $j=1$) and where the coefficient $c_j$ is either $\, (+1)\times$(the exponent in the $j$th entry)\, if that entry is a $0$, or \,$(-1)\times$(the exponent in the entry) if it is an $\infty$. Summing all these monomials, we obtain
\begin{equation}
\begin{matrix}
0 & \infty^k & 0\\[-3.75mm]\\
\downarrow\, & \downarrow\, & \downarrow\,\\[-3mm]\\
+1\cdot\lambda^0 & -k \cdot \lambda^1 & + 1\cdot \lambda^2
\end{matrix}\label{corresone}
\end{equation}
i.e. the polynomial $\lambda^2 - k \lambda +1$ for which we shall prove that it, in fact, determines the dynamical degree of the mapping (assuming that $k\in 2\mathbb{Z}_{>0}$). More precisely, when $k=2$, we obtain a polynomial  that does not have any roots greater than 1,  a fact that will be shown to  imply that the dynamical degree for the mapping in this case is exactly equal to 1 (as it should be, for a mapping that is a well-known member of the Quispel-Roberts-Thompson (QRT) family of integrable mappings \cite{qrt}). For $k\geq4$, even, the largest root of $\lambda^2 - k \lambda +1$ is
\begin{equation}
\lambda_* = \frac{k+\sqrt{k^2-4\,}}{2},
\label{dydegk}
\end{equation}
which is of course greater than 1 and, as we shall show in section \ref{secmapkprelims}, is nothing but the value of the dynamical degree for the mapping \eqref{mapk} for even values of $k$ (cf. \cite{nousPhysD} for a detailed analysis of the cases $k=2$ and $4$).

The aim of this paper is to show that this correspondence between the singularity patterns of a confining mapping and its dynamical degree is not a coincidence. Of course, when the mapping has more than one singularity pattern, and when these patterns involve more than two different values, this correspondence becomes more complicated. But as we shall show, it can still be easily established, through what is essentially our express method reformulated in algebro-geometric terms.

In the following we shall first explain the workings of the express method on two examples, the one above and on a simple multiplicative QRT mapping. We then give a brief overview of the geometric and growth-related properties of (confining) birational maps on $\cpp$  (section \ref{prelims}), after which we present a reformulation in algebro-geometric terms of the express method (as presented in \cite{rodone}) for three point mappings, i.e. for birational mappings that on $\cpp$ that are defined by a three point relation
\begin{equation}
x_{n+1}=F(x_{n-1},x_n),
\label{threepoint}
\end{equation}
and we prove the main theorems needed for this purpose (section \ref{mres}).
Finally, in section \ref{examps}, we explain on some examples how these results allow one to decide on the integrability of three point mappings, and how they can be used to obtain the dynamical degree, from nothing more than a simple analysis of their singularity patterns.

In the Appendix, we apply Halburd's method to obtain the exact degree sequence for the iterates of the two mappings of section \ref{secexpress} (Appendix 1) and we briefly explain how the degree sequence of the iterates for the non-confining cases of mapping \eqref{mapk} can be calculated (Appendix 2).

\section{The express method}\label{secexpress}
In this section we shall explain the method we proposed in \cite{rodone} for obtaining the dynamical degree of a confining mapping, directly from its singularity patterns, on two examples: mapping \eqref{mapk} from the introduction and a simple multiplicative QRT mapping which, when deautonomized, leads to a qPI equation. The full, Halburd-style, analysis of these mappings which yields the exact degrees of the iterates of the mapping, is given in Appendix 1.

\subsection{The express method for mapping \eqref{mapk}}\label{secmapk}
We consider the case $k\in 2\mathbb{Z}_{\geq1}$, for which mapping \eqref{mapk} has just one singularity, corresponding to the confined singularity pattern \eqref{sipak}: $(0\, \infty^k\, 0)$. 

In the express method one tries to estimate the number of pre-images of some value $\omega\in\cp$, for the nth iterate of a mapping viewed as a rational function $f_n(z)~(n\geq1)$ in the initial condition $x_1=z$. The other initial condition, $x_0$, is taken to be completely generic (i.e., it is supposed not to satisfy any specific relations) and is not considered to be a variable in $f_n(z)$, just an inconsequential parameter. For example, for mapping \eqref{mapk} one has $f_1(z):=x_1=z, f_2(z):=x_2=z^{-k}-x_0$, etc.
As in Halburd's method (cf. Appendix 1) we shall try to calculate the number of pre-images in $z$ of $\omega=f_n(z)$, for values $\omega$ that appear in the singularity pattern(s) for the mapping. However, in our case, we shall only count the number of pre-images approximately, as we are not interested in knowing the exact degree of $f_n(z)$ in $z$.

Let us first consider the pre-images of the value $0$. A value $0$ can of course appear `spontaneously' (as an accidental consequence of some choice of initial conditions) at some iterate $n$, but from the singularity pattern  \eqref{sipak} it is clear that such a value also {\it necessarily} arises 2 steps after a previous $0$.
If we denote the number of `spontaneous' appearances of 0 at the nth iterate of the mapping as $Z_n$, we find that the number of pre-images of 0 for $f_n(z)$ (when $n\geq3$) is given by $Z_n + Z_{n-2}$. Similarly, we also find that the number of pre-images of $\infty$ for $f_n(z)$ should be at least $k\times Z_{n-1}$, because a value $0$ in the singularity pattern \eqref{sipak} necessarily generates an $\infty$ with multiplicity $k$ at the next step. Neglecting any other possible occurrences of the value $\infty$, we then write
\begin{equation}
Z_n + Z_{n-2} \simeq k Z_{n-1}.
\label{Zrelk}
\end{equation}
Although this is only an approximate relation, not a genuine equality, we shall nonetheless say that it gives rise to a meaningful characteristic equation (by taking $Z_n\sim \lambda^n$):
\begin{equation}
\lambda^2 - k\lambda + 1 =0.
\label{chareqk}
\end{equation}
This is of course nothing but the characteristic equation for the polynomial obtained in the introduction, where we posited that the lack of (characteristic) roots greater than 1 when $k=2$ in fact tells us that the corresponding mapping is integrable (which it is) and that all other cases $k\in 2\mathbb{Z}_{\geq2}$ are in fact nonintegrable.

The main difference between our method and Halburd's method will become clear in Appendix 1, where it will be shown that the calculation of the exact degrees for this mapping actually requires more information than that which is contained in the singularity pattern \eqref{sipak}. The origin of this problem will also become clear in the following example.

{\remark That the greatest root of the equation \eqref{chareqk}, $\lambda_*=(k+\sqrt{k^2-4\,})/2$, is indeed the dynamical degree for \eqref{mapk} when $k \in 2\mathbb{Z}_{\geq1}$, will be shown in section \ref{secmapkprelims} using algebro-geometric tools, and in Appendix 1 by direct calculation of the degrees of the iterates of the mapping, using Halburd's method.}

{\remark When $k$ is an odd positive integer, mapping \eqref{mapk} does not enjoy the singularity confinement property and, as will be shown in Appendix 2, it is always nonintegrable when $k>1$. For $k=1$, however, the mapping is linearisable.}

\subsection{The express method for a multiplicative QRT map}\label{secqPI}
Let us analyse the singularities of the birational mapping given by the equation
\begin{equation}
x_{n+1} x_{n-1} = \frac{x_n+1}{x_n^2}.
\label{qPI}
\end{equation}
This mapping has three singularities (i.e., three values of $x_n\in\cp$ for which the next iterate, $x_{n+1}$, does not depend on $x_{n-1}$): $x_n=-1, 0$ and $\infty$.

Performing the singularity analysis, as sketched in the introduction, it is easy to verify that the singularity at $x_n=-1$ is, in fact, confined. The corresponding singularity pattern is
\begin{equation}
\sipa{-1~~ 0~~ \infty^2~~ 0~~ -1}.
\label{sipaqPI}
\end{equation}
The situation is slightly more complicated for the other two singularities as these turn out to be part of a cycle of length 8:
\begin{gather}
x_{n-1}=u, \quad x_{n}=0, \quad x_{n+1} = \infty^2, \quad x_{n+2} =0, \quad x_{n+3}=u', \quad x_{n+4}=\infty,  \nonumber\\
x_{n+5}=0, \quad x_{n+6} = \infty, \quad x_{n+7}=u, \quad x_{n+8}=0, \quad \cdots\,,
\label{cycleqPI}
\end{gather}
where $u'=1/u$. Within this cycle each singularity actually confines (after 3 steps, for both $0$ and $\infty$) but the cycle keeps repeating, indefinitely. We associate with this cycle the {\it cyclic} singularity pattern
\begin{equation}
\sipa{x_0~~ 0~~ \infty^2~~ 0~~ x_0'~~ \infty~~ 0~~ \infty},
\label{sipacyclqPI}
\end{equation}
and from here on, to emphasize the difference with the confined pattern \eqref{sipaqPI}, we shall refer to the latter as an {\it open} singularity pattern.

The difference between these two types of singularity patterns and the role each type has to play in our analysis will become clear in sections \ref{prelims} and \ref{mres}. The essence of the express method, however, lies in systematically neglecting any cyclic patterns, and to use only the open patterns to establish relations such as \eqref{Zrelk}, relations that will allow us to decide on the integrability or non-integrability of the mapping.

{\remark This last statement is actually a slight oversimplication as, for general three point mappings, there are still other types of singularity patterns to consider (cf. \cite{anticonf, review}). For confining mappings however, the above is an accurate characterization of the express method.}\medskip

In this spirit, we obtain the following relations from the open pattern \eqref{sipaqPI}:
\begin{equation}
M_{n-4}+M_n \simeq M_{n-1} + M_{n-3} \simeq 2 M_{n-2},
\label{MrelqPI}
\end{equation}
where $M_n$ denotes the number of spontaneous occurrences (pre-images) of $-1$ at the nth iterate of \eqref{qPI}, and where we have approximated the number of pre-images of the value 0 by $M_{n-1} + M_{n-3}$ and that of $\infty$ by $2 M_{n-2}$, purposely neglecting any contributions from the cyclic pattern \eqref{sipacyclqPI} for those values (cf. also Appendix 1).

Just as we did for the example in section \ref{secmapk}, we can now use any of the relations contained in \eqref{MrelqPI} to write a characteristic equation (taking $M_n\sim \lambda^n$):
\begin{gather}
\lambda^4-\lambda^3-\lambda+1 \equiv (\lambda^2+\lambda+1) (\lambda-1)^2=0\label{qPIchar1}\\
\lambda^4 - 2 \lambda^2 +1 \equiv (\lambda+1)^2 (\lambda-1)^2=0\label{qPIchar2}\\
\lambda^2-2\lambda+1 \equiv (\lambda-1)^2 = 0\,,\label{qPIchar3}
\end{gather}
all of which, however, lack a root greater than 1. As will be shown in section \ref{mres}, this observation allows us to conclude with certainty that the mapping \eqref{qPI} is integrable (which it indeed is, as it is a QRT mapping). 

Note that, in this case as well, there is a simple correspondence between the singularity pattern -- which now involves three different values -- and the above polynomials,
\begin{equation}
\begin{matrix}
-1 & 0 & \infty^2 & 0 & -1\\[-3.75mm]\\
\downarrow\, & \downarrow\, & \downarrow\,& \downarrow\, & \downarrow\,\\[-3mm]\\
& +1\cdot\lambda^0 & -2\cdot \lambda^1 & +1\cdot \lambda^2 & \\
+1\cdot\lambda^0 & -1\cdot \lambda^1 & & -1\cdot\lambda^3 & +1\cdot\lambda^4\\
+1\cdot\lambda^0 &   &  -2\cdot\lambda^2 &  & +1\cdot\lambda^4
\end{matrix}\label{correstwo}
\end{equation}
obtained by a straightforward generalization of the rule explained in the introduction, to any (ordered) pair of values ($0,\infty$), (-1,0) or (-1,$\infty$) that appear in the singularity pattern (from left to right). As we shall see in section \ref{mres}, this simple observation actually provides an important clue as to the interpretation of the relation \eqref{MrelqPI} in terms of the underlying geometry of the mapping \eqref{qPI}.

\section{Geometric properties of second order birational mappings}\label{prelims}

The idea that a second order birational mapping that only has confined singularities can always lifted to an automorphism on a rational surface, is already present (implicitly) in the work of Sakai \cite{sakai} and Takenawa \cite{takenawa-method} but was formulated explicitly for the first time by Takenawa in \cite{takenawaeguchi}. 
More precisely, a confining birational mapping on $\cpp$ always possesses a so-called {\it space of initial conditions}.
{\defi If for an autonomous birational map, $f: \cpp \dashrightarrow  \cpp$, there exists a rational surface $X$ and a birational map $g: X \dashrightarrow \cpp$ such that $\fie := g^{-1} \circ f \circ g$ is an automorphism on $X$,
$$\begin{matrix}
X &\changes{\xrightarrow[~~~^{^\sim} ~~]{~~~\fie~~}}& X \\
{\footnotesize\text{g}}\!\dashdownarrow{1.2}\,\,\, & & \,\,\,\dashdownarrow{1.2}\!{\footnotesize\text{g}} \\
\cpp & \xdashrightarrow{\text{~~~~f~~~}}& \cpp
\end{matrix}$$
then we call $X$ a `space of initial conditions' for $f$.
}
{\remark \changes{The birational map $g$ that, in the above sense, regularizes $f$ is in general a composition of a finite number of blow-ups and blow-downs. Note however that} the space of initial conditions for a confining map can always be constructed with a finite number of blow-ups only, i.e. there is no need for blow-downs in this case \cite{dillerfavre}.}

{\remark The notion of a space of initial conditions was extended to non-autonomous second order birational mappings in \cite{mase}. Strictly speaking, for a confining  nonautonomous mapping one does not have a single rational surface $X$, but rather a family of rational surfaces $X_n$ with a family of isomorphims \changes{$\fie_n: X_n\xrightarrow{\sim}{~} X_{n+1}$} acting between them. This extension of the notion of a space of initial conditions is non-trivial and involves a number of restrictions on the surfaces $X_n$ as well as on the birational transformations that give rise to them (see, in particular, Definition 2.5 and Remark 2.22 in \cite{mase}).}\medskip

Staying, for simplicity, with the autonomous case we shall denote by $\fie_*: \pik{X} \to \pik{X}$ the action induced by the automorphism $\fie$ on the Picard group  of the rational surface $X$. The Picard group $\pik{X}$ is a finitely generated free $\mathbb{Z}$-module and we shall denote its rank, the {\it Picard number} of the rational surface $X$ by the symbol $\rho$: $\rho := {\rm rank}\big(\pik{X}\big)$. The push-forward map $\fie_*$ can therefore be identified with an element of $GL_\rho(\mathbb{Z})$, which we shall denote $\Phi$.

As explained in \cite{takenawa-method} (and in a more general setting in \cite{dillerfavre}) the degree growth of the iterations of a confining mappping $f$ is governed by the spectrum of this matrix $\Phi$.
{\defi If we denote the nth iterate of a birational map $f:\cpp\dashrightarrow\cpp$ as $f^{(n)}$, then 
$$\lambda_* = \lim_{n\to+\infty} \big( \deg f^{(n)}\big)^{1/n}$$
is called the dynamical degree of the mapping $f$.
}
{\remark This limit always exists for autonomous mappings and it is a real number not less than 1. For nonautonomous mappings the limit need not exist in general, but it does exist if $f$ possesses a space of initial conditions in the sense of \cite{mase}.}\medskip

The following theorem holds for both autonomous as well as nonautonomous birational mappings on $\cpp$ (if the latter have a space of initial conditions, cf. \cite{mase}).

{\thm\label{phidyndeg}\cite{takenawa-method, dillerfavre}~ The dynamical degree of a confining birational mapping is given by the largest eigenvalue of the matrix $\Phi\in GL_\rho(\mathbb{Z})$ associated with the action of the mapping on the Picard group of its space of initial conditions.}\medskip

Diller and Favre have shown that the spectrum of such a regular matrix $\Phi$ is very special.

{\thm\label{DF} \cite{dillerfavre}~ Let $\Phi\in GL_\rho(\mathbb{Z})$ represent the action $\fie_*: \pik{X}\to\pik{X}$ on the Picard group for the space of initial conditions of a confining birational mapping on $\cpp$. The Jordan normal form of the matrix $\Phi$ can only take one of the following three forms:
\begin{itemize}
\item[{\rm (a)}]~ {\footnotesize $\begin{pmatrix}\nu_1\!\!\!\!&\!\!\!\!&\\\!\!\!\!&\ddots\!\!\!\!&\\\!\!\!\!&\!\!\!\!&\nu_\rho\,\end{pmatrix}$}, where all $\nu_j$ are roots of unity,
\item[{\rm (b)}]~ {\footnotesize $\begin{pmatrix}1\!\!\!\!& 1\!\!\!\!& 0\!\!\!\!& & & \\ 0\!\!\!\!& 1\!\!\!\!& 1\!\!\!\!& & & \\ 0\!\!\!\!& 0\!\!\!\!& 1\!\!\!\!& & & \\& & & \nu_1\!\!& & \\ & & & & \ddots\!\!& \\ & & & & &\!\!\nu_{\rho-3}\,\end{pmatrix}$}, where all $\nu_j$ are roots of unity,
\item[{\rm (c)}]~ {\footnotesize $\begin{pmatrix}\lambda\!\!\!\!& & & & \\ &1/\lambda\!\!\!\!& & & \\ & & \nu_1\!\!\!\!& & \\ & & & \ddots\!\!\!\!& \\ & & & & \nu_{\rho-2}\,\end{pmatrix}$}, where $\lambda>1$ and  $\vert\nu_j\vert=1 ~(j=1, \hdots, \rho-2)$.
\end{itemize}
}
{\remark\label{remkroneck}
Note that this shows that $\Phi$ can have at most one eigenvalue that is greater than 1, in which case this eigenvalue necessarily gives the value of the dynamical degree of the mapping (cf. Theorem \ref{phidyndeg}).
A similar result has been shown in \cite{mase} for the nonautonomous case. 
Moreover, it follows from a famous theorem by Kronecker \cite{kronecker}, that in the case (c) the eigenvalues with modulus 1 that are not roots of unity, are all Galois conjugates of the (unique) eigenvalue that is greater than 1.}\medskip

{\corol In case {\rm (a)} the birational mapping has bounded degree growth, in case {\rm (b)} the degree grows quadratically in $n$ and in case {\rm (c)} exponentially in $n$: $\deg f^{(n)} \sim \lambda^n$ (where $\lambda$ is the largest eigenvalue of the matrix, $\lambda>1$).}\medskip

For case {\rm (a)} the corresponding mapping is either periodic or birationally conjugate to a projective mapping on $\mathbb{P}^2$ \cite{blancdeserti} (in which case one might still call it integrable, although in a trivial sense). In the case {\rm (b)}, the dynamical degree of the mapping is equal to 1 and the mapping is integrable: in the autonomous case it has an invariant elliptic fibration \cite{dillerfavre}. In the case {\rm (c)} the dynamical degree is equal to $\lambda$ ($\lambda>1$) and the corresponding mapping is nonintegrable: in the autonomous case such a mapping does not have an invariant fibration \cite{dillerfavre} and it cannot have any nontrivial symmetries.

{\thm \cite{cantat-ann}~ If the dynamical degree of a birational map $f$ on $\cpp$ is greater than 1, then for any birational transformation $g$ on $\cpp$ that commutes with $f$ there exist two integers, $m\in\mathbb{Z}_{>0}$ and $n\in\mathbb{Z}$, such that $g^{(m)} = f^{(n)}$.}\medskip

Quite a lot is known about the number-theoretical properties of the dynamical degree.

{\thm \cite{dillerfavre}~ The dynamical degree for a confining, second order, birational mapping is 1, a reciprocal quadratic integer or a Salem number.}\medskip

A reciprocal quadratic integer is an integer with a minimal polynomial of the form $t^2 - a t +1$ for some $a\in\mathbb{Z}$ and a {\it Salem number} is an extension of this notion:

{\defi A Salem number is a real algebraic integer $\lambda>1$ such that $1/\lambda$ is a conjugate and all (but at least one) of its conjugates lie on the unit circle.}
Note that the minimal polynomial for a Salem number is necessarily palindromic and is at least \changes{of degree} 4.

As a matter of fact, the only second order birational mappings that have a Salem number as their dynamical degree are confining:
{\thm\cite{blanccantat}~If the dynamical degree for a  birational (autonomous) map $f:\cpp\dashrightarrow\cpp$ is a Salem number, then $f$ possesses as space of initial conditions.}\medskip

Of course, the Jordan normal form of the representation matrix $\Phi$ is, generally, inaccessible when working over $\pik{X}$ (as its construction involves factorization over $\mathbb{C}$). However, allowing changes of basis over
\begin{equation}
\pikq{X}:= \bigoplus_{j=1}^\rho \mathbb{Q} F_j,
\end{equation}
for example, where $F_1, \hdots, F_\rho$ are a ($\mathbb{Z}$) basis for $\pik{X}$, it is possible to put $\Phi$ in rational (Frobenius) normal  form. Here, we shall content ourselves with a form of $\Phi$ that is close to (but not quite the same as) its Frobenius form. In fact, one can always find a basis for $\pik{X}$ that yields a matrix representation of $\fie_*: \pik{X} \to \pik{X}$ of the form
\begin{equation}
\begin{pmatrix}
U & {\large \ast} \\
{\large 0} & \!A
\end{pmatrix},
\label{phibf}
\end{equation}
where $U$ is a unitary matrix with integer entries (or, in other words, a {\it signed permutation matrix}) of size $p$ and where $A\in GL_{\rho-p}(\mathbb{Z})$, for some positive integer $p$. The symbol {\large$0$} represents a zero sub-matrix of size $(\rho-p)\times p$ and ${\large\ast}$ an element in $M_{p\times(\rho-p)}(\mathbb{Z})$ which will not play any role in our analysis.

Just as when transforming a matrix to its Frobenius normal form, the form \eqref{phibf} requires a change of basis involving cyclic (Krylov) subspaces, but in our case we do not aim at a full decomposition of $\pik{X}$ into cyclic subspaces. In fact, the $A$-block in \eqref{phibf} corresponds to the action of $\fie_*$ induced on the quotient $\pik{X}/\PP{X}$, which we shall denote as $\pfie$,
\begin{equation}
\pfie: \pik{X}/\PP{X}\to \pik{X}/\PP{X},\quad \pfie(F) = \changes{\fie_*(F)}\mod \PP{X},
\label{pushforward}
\end{equation}
where $\PP{X}$ is the {\it cyclic part} of the Picard group:
\begin{equation}
\PP{X} :=\left\{ F\in\pik{X}\,\big|~ ^\exists m\in\mathbb{Z} : ~ \fie_*^m F = F \right\}.
\label{ppx}
\end{equation}

{\remark One can always find a $\mathbb{Z}$-basis for $\pik{X}$ such that \changes{$\fie_*$} is represented by a matrix of the form \eqref{phibf} \changes{since} both $\PP{X}$ and $\pik{X}/\PP{X}$ are free $\mathbb{Z}$-modules. This last statement follows from the fact that $\pik{X}/\PP{X}$ is torsion free because $\PP{X}$ is a saturated $\mathbb{Z}$-submodule of $\pik{X}$, i.e.: if for some $F\in\pik{X}$ there exists a non-zero integer $\ell$ such that $\ell F\in\PP{X}$, then $F\in\PP{X}$.}\medskip

We then have the following obvious corollary from Theorem \ref{DF}. 
{\corol \label{corol2}~Let us denote the minimal polynomials for $\fie_*$ and $\pfie$ as $\mu_{\fie_*}(t)$ and $\mu_{\pfie}(t)$, respectively. 
\par\noindent
\begin{itemize}
\item[--] In case {\rm (a)} of Theorem \ref{DF}, $\mu_{\fie_*}(t)$ is a product of cyclotomic polynomials, all different. Moreover $\PP{X}=\pik{X}$, i.e. $\pik{X}/\PP{X}=0$.
\item[--] In case {\rm (b)}, $\mu_{\fie_*}(t) = (t-1)^3 \times$(a product of cyclotomic polynomials in $t$, all different and different from $(t-1)$). Moreover, $\mu_{\pfie}(t) = (t-1)^2$.
\item[--] In case {\rm(c)}, $\mu_{\fie_*}(t) = \mu_\lambda(t) \times$(a product of cyclotomic polynomials, all different), where $\mu_\lambda(t)$ is the minimal polynomial for the quadratic integer or Salem number $\lambda$ that gives the largest eigenvalue for the matrix $\Phi$ (i.e., the dynamical degree of the mapping). Moreover, $\mu_{\pfie}(t) = \mu_\lambda(t)$ (cf. Remark \ref{remkroneck}).
\end{itemize}
}

{\remark\label{remgauss} Since any polynomial with integer coefficients that is \changes{irreducible over the integers is also irreducible over the rationals} (due to a famous Lemma by Gauss), the polynomial $\mu_{\pfie}(t)$ is irreducible over $\mathbb{Q}[t]$ and is therefore indeed the minimal polynomial of the dynamical degree.}\medskip

Casting aside the case (a) with bounded degree growth, it is clear that it is the minimal polynomial $\mu_{\pfie}(t)$ for the submatrix $A$ in \eqref{phibf} that decides on the integrability of the mapping. For example, if one can ascertain that $\mu_{\pfie}(t)$ contains a factor $(t-1)$ then one is certain that the mapping is integrable (with quadratic degree growth), whereas if $\mu_{\pfie}(t)$ has a root that is greater than 1 then the mapping has exponential degree growth and is therefore nonintegrable.

In the next section we will show that there exists in fact a very simple procedure for answering this question, a procedure which is tantamount to the express method, though purely geometric in nature. Let us first, however, illustrate the above definitions and results on the two mappings we used in section \ref{secmapk}. This will also serve to highlight the (not quite unexpected) link between the cyclic singularity patterns we introduced in the previous sections, and the subgroup $\PP{X}$ \eqref{ppx} defined above.

\subsection{The geometric structure of mapping \eqref{qPI}}\label{secqPIprelims}
As a first example we take the mapping \eqref{qPI} from section \ref{secqPI}, which we write as a mapping on $\cpp$ as:
\begin{equation}
f : \cpp \dashrightarrow \cpp\,,\quad \ve{x}{y} \mapsto \ve{y}{(y+1)/(xy^2)}.
\label{qPIcpp}
\end{equation}
The open singularity pattern \eqref{sipaqPI} then corresponds to the following iteration of the mapping in which the curve $[y=-1]$ in $\cpp$ collapses to a point which, after 4 more iterations of the mapping, again gives rise to a curve:
\begin{equation}
[y=-1]\xrightarrow{~f} \ve{-1}{0}\to\ve{0}{\infty^2}\to\ve{\infty^2}{0}\to\ve{0}{-1}\to [x=-1] .
\label{qPIcppopen}
\end{equation}
The cyclic singularity pattern \eqref{sipacyclqPI} corresponds to
\begin{equation}
\hskip-.25cm[y=0] \xrightarrow{~f} \ve{0}{\infty^2}\to\ve{\infty^2}{0}\to [x=0] \to [y=\infty] \to \ve{\infty}{0}\to \ve{0}{\infty} \to [x=\infty] \to [y=0] \to \cdots
\label{qPIcppcycle}
\end{equation}

{\remark\label{zenotation} \changes{The notation {\footnotesize$\ve{0}{\infty^2}$} we use here (and, mutatis mutandis, {\footnotesize$\ve{\infty^2}{0}$}) should be understood as follows. The $\infty^2$ symbol is that introduced in Section \ref{intro} and which was already used in the singularity patterns \eqref{sipaqPI} and \eqref{sipacyclqPI}. If we interpret {\footnotesize$\ve{0}{\infty^2}$} as a point in $\cpp$ then it is identical to the point {\footnotesize$\ve{0}{\infty}$} but our notation offers extra information: as we shall see, after regularising the mapping by blow-up, {\footnotesize$\ve{0}{\infty^2}$} should be thought of as representing a curve on a rational surface and the exponent 2 in $\infty^2$ then refers to the multiplicity of the total transform of the divisor of $[y=\infty]$ in $\cpp$ with respect to the divisor for that curve.
}}\medskip

\begin{figure}[t]
\begin{center}
\resizebox{6.25cm}{!}{\includegraphics{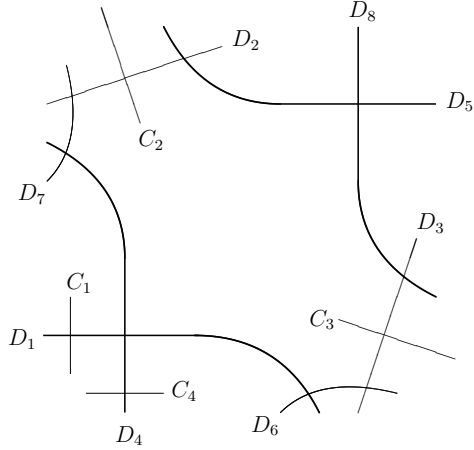}}
\caption{Schematic representation of the space of initial conditions for the mapping \eqref{qPIcpp}. }
\label{XqPI}
\end{center}
\end{figure}

\changes{
From the above two singularity patterns it is easy to ascertain that to regularise this mapping one needs at least 8 blow-ups: one at $\footnotesize\ve{-1}{0}$ and $\footnotesize\ve{0}{-1}$ each, and three each at $\footnotesize\ve{0}{\infty}$ and $\footnotesize\ve{\infty}{0}$.
}
In fact, these blow-ups turn out to be sufficient. Skipping over the details (of what amounts to a standard calculation), after blow-up, the mapping is easily seen to correspond to an automorphism $\fie$ of the rational surface $X$ depicted in Figure \ref{XqPI}. 
\changes{The curves $C_1, C_2, C_3$ and $C_4$ are the exceptional curves (i.e. with self-intersection -1) of the (last) blow-ups at $\footnotesize\ve{-1}{0}$, $\footnotesize\ve{0}{\infty}$, $\footnotesize\ve{\infty}{0}$ and $\footnotesize\ve{0}{-1}$ respectively. }
The curves $D_1, D_4, D_5$ and $D_8$ are the strict transforms of the curves $[y=0], [x=0], [y=\infty]$ and $[x=\infty]$ on $\cpp$, and $D_7$ and $D_2$ ($D_6$ and $D_3$) are the results of consecutive blow-ups at $\footnotesize\ve{0}{\infty}$ (and $\footnotesize\ve{\infty}{0}$), in that order. All curves labelled $D_j$ have self-intersection -2.
\changes{The meaning of the exponent 2 in $\infty^2$ in our notation can now be understood from the expression for the total transform of the divisor of $[y=\infty]$, which is (cf. Figure \ref{XqPI}) 
\begin{equation}
D_5+D_7 + 2 (D_2+C_2),
\end{equation}
in which $C_2$ and $D_2$  are the curves that correspond to the {\footnotesize$\ve{0}{\infty^2}$} in the open pattern \eqref{qPIcppopen} and in the cyclic pattern \eqref{qPIcppcycle}, respectively.
}

The action of the automorphism $\fie: X\to X$ on these curves can of course be read off directly from the patterns \eqref{qPIcppopen} and \eqref{qPIcppcycle}:
\begin{gather}
D_1 \to D_2 \to \cdots \to D_7\to D_8 \to D_1 ,\label{qPIcyclic}\\
\{y=-1\}\to C_1 \to C_2 \to C_3 \to C_4 \to \{x=-1\},\label{qPIopen}
\end{gather}
where by $\{y=-1\}$ and $\{x=-1\}$ we denote the strict transforms of the curves $[y=-1]$ and $[x=-1]$ in $\cpp$.

{\remark Although these are no longer `singularity' patterns as all singularities have been resolved, the analogy with the original singularity patterns is perfect and we shall refer to \eqref{qPIcyclic} as a cyclic pattern and to \eqref{qPIopen} as an open pattern for the automorphism $\fie$ on $X$.}\medskip

Taking (the divisor class for each one of) the curves $D_1, \hdots, D_8, C_1$ and $C_2$ as a basis for $\pik{X}$ for this surface (which has Picard number $\rho=10$) we obtain the following matrix representation for the push-forward map $\fie_*:\pik{X}\to\pik{X}$:
\begin{equation}
\Phi=
{\footnotesize
\begin{pmatrix}
\,0\!\!&\!\!0\!\!&\!\!0\!\!&\!\!0\!\!&\!\!0\!\!&\!\!0\!\!&\!\!0\!\!&\!\!1\!\!&\!\!0\!\!&\!\!-1\\
1\!\!&\!\!0\!\!&\!\!0\!\!&\!\!0\!\!&\!\!0\!\!&\!\!0\!\!&\!\!0\!\!&\!\!0\!\!&\!\!0\!\!&\!\!2\\
0\!\!&\!\!1\!\!&\!\!0\!\!&\!\!0\!\!&\!\!0\!\!&\!\!0\!\!&\!\!0\!\!&\!\!0\!\!&\!\!0\!\!&\!\!-1\\
0\!\!&\!\!0\!\!&\!\!1\!\!&\!\!0\!\!&\!\!0\!\!&\!\!0\!\!&\!\!0\!\!&\!\!0\!\!&\!\!0\!\!&\!\!0\\
0\!\!&\!\!0\!\!&\!\!0\!\!&\!\!1\!\!&\!\!0\!\!&\!\!0\!\!&\!\!0\!\!&\!\!0\!\!&\!\!0\!\!&\!\!1\\
0\!\!&\!\!0\!\!&\!\!0\!\!&\!\!0\!\!&\!\!1\!\!&\!\!0\!\!&\!\!0\!\!&\!\!0\!\!&\!\!0\!\!&\!\!-1\\
0\!\!&\!\!0\!\!&\!\!0\!\!&\!\!0\!\!&\!\!0\!\!&\!\!1\!\!&\!\!0\!\!&\!\!0\!\!&\!\!0\!\!&\!\!1\\
0\!\!&\!\!0\!\!&\!\!0\!\!&\!\!0\!\!&\!\!0\!\!&\!\!0\!\!&\!\!1\!\!&\!\!0\!\!&\!\!0\!\!&\!\!0\\
0\!\!&\!\!0\!\!&\!\!0\!\!&\!\!0\!\!&\!\!0\!\!&\!\!0\!\!&\!\!0\!\!&\!\!0\!\!&\!\!0\!\!&\!\!-1\\
0\!\!&\!\!0\!\!&\!\!0\!\!&\!\!0\!\!&\!\!0\!\!&\!\!0\!\!&\!\!0\!\!&\!\!0\!\!&\!\!1\!\!&\!\!2\,
\end{pmatrix}
}
\label{qPIPhi}
\end{equation}
The last column in this matrix is obtained by expressing the linear equivalence, on $\pik{X}$, of the total transforms of the \changes{divisors of the} curves $[y=0]$ and $[y=\infty]$:
\begin{gather}
D_1 + C_1 + D_6 + D_3 + C_3 \changes{\,\sim\,} D_5 + D_7 + 2 (D_2 + C_2) \\
\Leftrightarrow\quad \fie_*\,   C_2 = C_3 \changes{\,\sim\,} -D_1 + 2 D_2 - D_3 + D_5 - D_6 + D_7 - C_1 + 2 C_2,
\end{gather}
\changes{where the symbol $\sim$ denotes linear equivalence as divisors on $X$.
{\remark\label{samesymbol} Hereafter, we shall denote curves on the surface $X$ and their representatives in $\pik{X}$ by the same symbol. As the context will always be clear however, we do not think this will lead to any confusion. Moreover, we shall also write a simple equality $=$ instead of $\sim$ for expressing linear equivalence.}
}\medskip

Comparing the matrix \eqref{qPIPhi} with the form \eqref{phibf} it is clear that in this case we have a size 8 permutation matrix $U$ that corresponds to the cyclic pattern \eqref{qPIcyclic}, whereas the submatrix $A$ is given by
\begin{equation}
A=\begin{pmatrix} 0 & -1 \\ 1 & 2\end{pmatrix}.
\end{equation}
Clearly, in this case, the cyclic part of the Picard group $\PP{X}$ is spanned by $D_1, \hdots, D_8$.

It is not difficult to check that the minimal polynomial for $\Phi$ is
\begin{equation}
\mu_{\fie_*}(t) = (t+1) (t^4+1) (t^2+1) (t-1)^3,
\end{equation}
the form of which matches exactly that predicted by Corollary \ref{corol2} for a mapping with quadratic degree growth. The minimal polynomial for $A$ is of course 
\begin{equation}
\mu_{\pfie}(t) = (t-1)^2,
\end{equation}
just as predicted by Corollary \ref{corol2}.

\subsection{The geometric structure of mapping \eqref{mapk}}\label{secmapkprelims}
We first write equation \eqref{mapk} as a birational mapping on $\cpp$:
\begin{equation}
f : \cpp \dashrightarrow \cpp\,,\quad \ve{x}{y} \mapsto \ve{y}{-x + y^{-k}},
\label{mapkcpp}
\end{equation}
where $k$ is a positive even integer.

\begin{figure}[t]
\begin{center}
\resizebox{8.5cm}{!}{\includegraphics{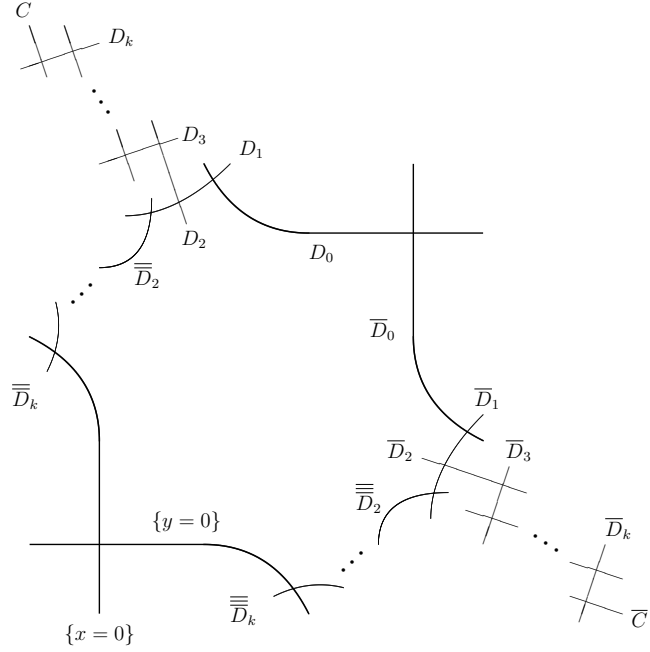}}
\caption{Schematic representation of the space of initial conditions for the mapping \eqref{mapkcpp}. }
\label{Xmapk}
\end{center}
\end{figure}

The confined singularity pattern \eqref{sipak} corresponds to the chain
\begin{equation}
[y=0] \xrightarrow{~f} \ve{0}{\infty^k}\to\ve{\infty^k}{0}\to [x=0].
\end{equation}
Again skipping the details of the blow-ups, \changes{it turns out that the mapping becomes an automorphism $\fie$ on the rational surface $X$ depicted in Figure \ref{Xmapk}, after $4k$ blow-ups: $2k$ successive blow-ups at the point $\footnotesize\ve{0}{\infty}$, giving rise rise to the curves $\overline{\overline{D}}_k, \cdots,  \overline{\overline{D}}_2, D_1, D_2, D_3, \cdots D_k$ and $C$ (in that order) and $2k$ blow-ups at $\footnotesize\ve{\infty}{0}$ giving rise to the curves 
$\overline{\overline{\overline{D}}}_k,  \cdots, \overline{\overline{\overline{D}}}_2,  \overline{D}_1,  \overline{D}_2,  \overline{D}_3, \cdots  \overline{D}_k$ and $\overline{C}$.
}

The action of the automorphism $\fie$ on the curves labelled $D_j$ (with and without bars) takes the form of the following $(k+1)$ cyclic patterns:
\begin{gather}
D_0\to\overline{D}_0\to D_0 ,\qquad D_1\to\overline{D}_1\to D_1\\
D_j\to\overline{D}_j\to\overline{\overline{D}}_j\to\overline{\overline{\overline{D}}}_j\to D_j\qquad (j=2, \cdots, k),\label{mapkfourcycles}
\end{gather} 
where the curves $D_1, \overline{D}_1$ and $D_j, \overline{D}_j, \overline{\overline{D}}_j, \overline{\overline{\overline{D}}}_j~(j=2, \cdots, k)$ are -2 curves and $D_0, \overline{D}_0$, the strict transforms of the curves $[y=\infty], [x=\infty]\subset\cpp$, are $-k$ curves. The curves $C$ and $\overline{C}$ are $-1$ curves that \changes{constitute} the open pattern:
\begin{equation}
\{y=0\}\to C \to \overline{C} \to \{x=0\},
\end{equation}
where $\{y=0\}$ and $\{x=0\}$ denote the strict transforms of the curves $[y=0]$ and $[x=0]$ in $\cpp$.

The exchange of the curves $D_0$ and $ \overline{D}_0$, 
\begin{equation}
\fie(D_0)=\overline{D}_0, \qquad\fie(\overline{D}_0)=D_0, 
\end{equation}
is of course the equivalent, on the rational surface $X$ depicted in Figure \ref{Xmapk}, of the (non-singular) cyclic pattern \eqref{mapkcycle} for  $[y=\infty]$ and $[x=\infty]$ in $\cpp$ that we describe in Appendix 1. 

If we take \changes{the divisor classes for} $D_0, \overline{D}_0,D_1, \overline{D}_1, D_2, \overline{D}_2, \overline{\overline{D}}_2, \overline{\overline{\overline{D}}}_2, \cdots, D_k, \overline{D}_k, \overline{\overline{D}}_k, \overline{\overline{\overline{D}}}_k, C$ and $\overline{C}$ as a basis for the Picard group $\pik{X}$ for this surface (which has Picard number $\rho=4k+2$) we obtain the following matrix representation for $\fie_*: \pik{X}\to\pik{X}$:\changes{
\begin{equation}
\Phi={\footnotesize\begin{pmatrix}
U_2\!\!&\!\!\!\!&\!\!\!\!&\!\!\!\!&\!\!\!\!&{\,*}\\
\!\!&\!\!U_2\!\!&\!\!\!\!&\!\!\!\!&\!\!\!\!&{\,*}\\
\!\!&\!\!\!\!&\!\!U_4\!\!&\!\!\!\!&\!\!\!\!&{\, *}\\
\!\!&\!\!\!\!&\!\!\!\!&\!\!\ddots\!\!&\!\!\!\!&\vdots\\
\!\!&\!\!\!\!&\!\!\!\!&\!\!\!\!&\!\!U_4\!\!&{\, *}\\
\!\!&\!\!\!\!&\!\!\!\!&\!\!\!\!&\!\!\!\!&\!\!A\
\end{pmatrix}},
\label{Phimapk}
\end{equation} }
where the submatrices $U_2, U_4$ and $A$ have the form
\begin{equation}
U_2=\begin{pmatrix}0&1\\1&0\end{pmatrix} ,\qquad U_4=\begin{pmatrix}0&0&0&1\\1&0&0&0\\0&1&0&0\\0&0&1&0\end{pmatrix} ,\qquad A=\begin{pmatrix}0&-1\\1&k\end{pmatrix},
\end{equation}
and where all other entries in $\Phi$ are zero, except for those in the last column (i.e. the $4k+2$nd column). 

The entries in this last column, i.e. the expression for $ \{x=0\}=\fie_*\, \overline{C}$ in the basis we have chosen, can be obtained by explicitly calculating the linear equivalence of the total transforms of the curves $[x=0], [x=\infty]\subset\cpp$ after blow-up:
 \begin{gather}
 \{x=0\} + \overline{\overline{D}}_k + \cdots + \overline{\overline{D}}_2 + D_1 +D_2 + \cdots + D_k + C\hskip5cm\nonumber \\
 \hskip3cm = \overline{D}_0 + k (\overline{D}_1 + \cdots + \overline{D}_k + \overline{C}) + (k-1) \overline{\overline{\overline{D}}}_2 + (k-2) \overline{\overline{\overline{D}}}_3+ \cdots + 1\cdot \overline{\overline{\overline{D}}}_k,
 \label{lineqmapk}
  \end{gather}
 which yields
 \begin{equation}
 \{x=0\} = {\cal D} - C + k \overline{C},
 \end{equation}
 for some element ${\cal D}\in\PP{X}$ determined by \eqref{lineqmapk}, where $\PP{X}$ is obviously spanned by $D_0, \overline{D}_0,D_1, \overline{D}_1, D_2, \overline{D}_2,$ $\overline{\overline{D}}_2, \overline{\overline{\overline{D}}}_2, \cdots, D_k, \overline{D}_k, \overline{\overline{D}}_k$ and $\overline{\overline{\overline{D}}}_k$.

The minimal polynomial of $\Phi$ is found to be
\begin{equation}
\mu_{\fie_*}(t) = (t+1) (t^2+1) (t-1) (t^2-k t+1),
\end{equation}
whereas the minimal polynomial of $\pfie$ (i.e., that of the submatrix $A$) is of course
\begin{equation}
\mu_{\pfie}(t) = (t^2-k t+1),
\end{equation}
exactly as in Corollary \ref{corol2}. The largest root of $\mu_{\pfie}(t)$ (and of course of $\mu_{\fie_*}(t)$), $\lambda_*=(k + \sqrt{k^2-4})/2$, is exactly 1 when $k=2$ (which therefore corresponds to a mapping with quadratic degree growth) and is greater than 1 when $k$ is an even integer greater than $2$ (a nonintegrable case of this mapping).

\section{The express method revisited}\label{mres}

We have seen in the previous section how the dynamical degree of a confining mapping is, in actual fact, determined by the action induced by the mapping on the non-cyclic part of the Picard group for the space of initial conditions for the mapping. In this section we will show that this induced action can, in a certain way, be characterized by merely analysing the open singularity patterns for the mapping, without actually constructing the space of initial conditions.

We shall only be concerned with mappings that are defined by three point relations such as \eqref{threepoint}, which we invariably take to define a birational mapping $f: \cpp\dashrightarrow\cpp$ of the form
\begin{equation}
\ve{x}{y}\mapsto\ve{y}{F(x,y)},
\label{mapCFcpp}
\end{equation}
such that the singularity patterns of the mapping can always be expressed in the $y$-coordinate only.

From here on, we shall exclude mappings that fall into the category (a) of Theorem \ref{DF}, i.e. mappings with bounded degree growth, for which the cyclic part of the Picard group, $\PP{X}$ (as defined as in \eqref{ppx}), is in fact the entire Picard group, i.e.: $\pik{X}/\PP{X}=0$.
Furthermore, we shall focus on the action $\pfie$ induced by $\fie_*$ on the non-cyclic part of $\pik{X}$, $\pfie: \pik{X}/\PP{X}\to\pik{X}/\PP{X}$, as defined in \eqref{pushforward}.

In the previous section we have also seen that the open singularity patterns for the original mapping, on $\cpp$, give rise to open patterns of curves on the space of initial conditions $X$, obtained by blowing-up the points that appear in the singularity patterns. As those patterns are by definition non-cyclic, we have the following Lemma.

{\lemma\label{trivlem} If $C\subset X$ is a curve that appears in an open pattern on $X$ (the equivalent of an open singularity pattern for the mapping, after blow-up), then its corresponding divisor class does not lie in the cyclic part of the Picard group, i.e.: $C\notin \PP{X}$ (following the notational convention set out  in Remark \ref{samesymbol}).

On the other hand, if $C\subset X$ is a curve that appears in a cyclic pattern corresponding to a cyclic singularity pattern for the mapping or, more generally, if 
$C\subset X$ is a curve that arose in the blow-up procedure but which is not part of any open pattern on $X$, then $C\in\PP{X}$.}

{\remark That the divisor class of a curve \changes{$C$} in an open pattern on $X$ cannot lie in $\PP{X}$ follows from the fact that (by definition) \changes{$C$} is not part of a cyclic pattern and because such a curve must have negative self-intersection, which implies that \changes{its divisor class} cannot be expressed as a non-trivial sum of effective classes in $\pik{X}$.}\medskip

We then have the following simple, but crucial, proposition.
{\prop\label{prop1} Suppose we have a polynomial $\psi(t)\in\mathbb{Q}[t]\setminus\{0\}$ and a \changes{class} $F\in\pik{X}/\PP{X}, F\neq0$, such that 
$$\psi(\fie_*)\, F = 0 \mod \PP{X}.$$
\begin{itemize}
\item[i)] If $\psi(t)$ does not have a root greater (or less) than 1, then we are in the case (b) of Theorem \ref{DF} and the mapping 
has a dynamical degree equal to 1 (and is integrable with quadratic degree growth). 
\item[ii)] If $\psi(1)\neq0$, then we are in the case (c) of Theorem \ref{DF} and the mapping has exponential degree growth (and is nonintegrable).
\end{itemize}
}

{\it Proof:} First of all, for any endomorphism $f$ of a finite dimensional vector space $V$ over some field $K$, it holds that if there exists a polynomial $\psi(t)\in  K[t]\setminus\{0\}$ such that $\psi(f) v =0$ for some non-zero $v\in V$, then $\psi(t)$ and the minimal polynomial for $f$ share a common factor over $K[t]$.

Hence, if $\psi(t)\in \mathbb{Q}[t]\setminus\{0\}$ and $F\in\pik{X}/\PP{X}, F\neq0 \mod \PP{X}$, such that $\psi(\fie_*)\, F = 0 \mod \PP{X}$, then $\psi(t)$ and $\mu_{\pfie}(t)$  (as defined in Corollary \ref{corol2}) must have a common factor. Since we excluded the case (a), from Corollary \ref{corol2} it then follows that if $\psi(t)$ does not have a root greater (or less) than 1, we cannot be in the case (c) and therefore must be in the case (b). Conversely, if $\psi(1)\neq0$ we must be in the case (c) since $\mu_{\pfie}(t)=(t-1)^2$ in the case (b).\hfill{$\blacksquare$}

{\example\label{ex41} ~\!\!}For mapping \eqref{mapkcpp} of section \ref{secmapkprelims}, we have just one very simple (open) singularity pattern, $(0~~\infty^k~~0)$, where $k$ is a positive even integer. We know that, after blow-up, the mapping has \changes{the space of initial conditions $X$ depicted in Figure \ref{Xmapk},} on which there will be curves $C$ and $\overline{C}$ that form the open pattern 
\begin{equation}
\{y=0\}\to C \to \overline{C} \to \{x=0\},
\end{equation}
which is the exact analogue of the open singularity pattern $(0^1~\infty^k~0^1)$. 

Since $C$
 and $\overline{C}$ are part of an open pattern on $X$, we have $C, \overline{C}\notin \PP{X}$ because of Lemma \ref{trivlem}, and thus $\fie_*\, C = \overline{C} \notin {\rm span}\big(C\big) \mod \PP{X}$, but $\fie_*\, \overline{C}\in {\rm span}\big( C, \overline{C}\,\big) \mod\PP{X}$. Denoting \changes{the class of the total transform of the curve} $[y=0]\subset\cpp$ as $H_y\in\pik{X}$, we have  that 
\begin{equation}
 H_y = \{y=0\} + \overline{C} \mod \PP{X} \equiv (\fie_*^{-1} + \fie_*)\, C \mod \PP{X},
\end{equation}
\changes{where $\overline{C}$ corresponds to $\footnotesize\ve{\infty^k}{0}$ (in the sense explained in Remark \ref{zenotation}). As before, $\{y=0\}$ is the \changes{class of the} strict transform of $[y=0]\subset\cpp$.} On the other hand, by linear equivalence on $\pik{X}$, we also have that
\begin{equation}
H_y= k C \mod\PP{X},
\end{equation}
\changes{where $C$ corresponds} to $\footnotesize\ve{0}{\infty^k}$ \changes{(as in Remark \ref{zenotation})}, hence the multiplicity $k$. Note that \changes{the class of} $\{y=\infty\}$, the strict transform of $[y=\infty]\subset\cpp$,  lies in $\PP{X}$ because it belongs to a cyclic pattern (Lemma \ref{trivlem}). We thus find that $k C =  (\fie_*^{-1} + \fie_*)\, C \mod \PP{X}$ or,
\begin{equation}
\big( \fie_*^2 - k \fie_* +1 \big)\, C = 0 \mod \PP{X},
\label{mapkmodP}
\end{equation}
for $C\neq 0 \mod\PP{X}$. When $k=2$, as the polynomial that annihilates $C$ in \eqref{mapkmodP} does not have any roots different form 1, we conclude from Proposition \ref{prop1} that the dynamical degree of the mapping in this case must be 1. If $k>2$ however, the corresponding polynomial does not have a root equal to 1 and we conclude, again from Proposition \eqref{prop1}, that the mapping has exponential degree growth.
 
 {\remark Since the polynomial $(t^2 - kt +1)$ is irreducible over $\mathbb{Z}$ for $k>2$,  we can conclude that it is nothing but the minimal polynomial for the dynamical degree for the mapping when $k\in2\mathbb{Z}_{>1}$ (cf. Remark \ref{remgauss}). Note that, in this way, we have found the (unique) eigenvalue for $\fie_*$ greater than 1, without having to calculate the entire matrix \eqref{Phimapk}.}
 
 {\example\label{ex42}~\!\!} For mapping \eqref{qPIcpp} of section \ref{secqPIprelims} we have the open singularity pattern: $(-1~~0~~\infty^2~~0~~-1)$, which is more complicated than that for Example \ref{ex41}. The corresponding open pattern on $X$, the space of initial conditions for this mapping, takes the form
\begin{equation}
\{y=-1\}\to C_1 \to C_2 \to C_3 \to C_4 \to \{x=-1\},
\end{equation} 
where $C_1, C_2, C_3$ and $C_4$ \changes{correspond to $\footnotesize\ve{-1}{0}, \ve{0}{\infty^2},\ve{\infty^2}{0}$ and $\footnotesize\ve{0}{-1}$, respectively (cf. Figure \ref{XqPI})}. Hence $C_j\neq0 \mod\PP{X}~(\,^\forall j=1, \hdots,4)$, due to Lemma \ref{trivlem}. From this lemma we also know that besides $\{y=-1\}$ and $\{x=-1\}$ (\changes{the classes of} the strict transforms of $[y=-1]$ and $[x=-1]$, respectively), all other \changes{classes of} curves that arise during \changes{the} blowing-up necessarily lie in $\PP{X}$. Moreover, denoting the \changes{class of the} total transform of $[y=0]\subset\cpp$ as $H_y\in\pik{X}$, we have\changes{
\begin{equation}
H_y=C_1 + C_3  \mod\PP{X} \equiv \big(\fie_*^2 + 1\big)\, C_1 \mod\PP{X},\label{qPIHy0}
\end{equation}
where we have used the fact that $\{y=0\}\in\PP{X}$, and also, by linear equivalence on $\pik{X}$
\begin{equation}
H_y=\{y=-1\} + C_4  \mod\PP{X} \equiv \big(\fie_*^{-1} + \fie_*^3\big)\, C_1 \mod\PP{X},\label{qPIHym1}
\end{equation}
}or
\begin{equation}
H_y= 2 C_2  \mod\PP{X} \equiv 2\fie_*\, C_1 \mod\PP{X},\label{qPIHyinf}
\end{equation}
where we have used the fact that $\{y=\infty\}\in\PP{X}$. Combining these expressions we find 3 non-trivial polynomials that annihilate $C_1$ on $\pik{X}/\PP{X}$, 
\begin{gather}
\big( \fie_*^4 - \fie_*^3 - \fie_* +1\big)\, C_1 = 0 \mod\PP{X}\label{qPIann1}\\
\big( \fie_*^4 -2\fie_*^2 +1\big)\, C_1 = 0 \mod\PP{X}\label{qPIann2}\\
\big( \fie_*^2 -2\fie_* +1\big)\, C_1 = 0 \mod\PP{X}\label{qPIann3}
\end{gather}
None of the above polynomials possess a root greater (or less) than 1 and from Proposition \ref{prop1} we can therefore conclude that the mapping must be integrable. Note that we were able to reach this conclusion without performing any actual blow-ups or without having access to the entire matrix \eqref{qPIPhi} for $\fie_*$.

{\remark The above expressions \eqref{qPIann1}, \eqref{qPIann2} and \eqref{qPIann2} are in one to one correspondence with the characteristic polynomials \eqref{qPIchar1}, \eqref{qPIchar2} and \eqref{qPIchar3} obtained from the express method for this mapping. This is not surprising as the original relation \eqref{MrelqPI} in the express method obviously corresponds to the action of the pullback $\fie^*$, which is related to the push-forward by $\fie_*=(\fie^{-1})^*$, on the curves $C_j$. Moreover, the construction of the equations (\ref{qPIann1}$\sim$\ref{qPIann2}) itself also corresponds perfectly to the direct construction of the characteristic polynomials \eqref{qPIchar1}, \eqref{qPIchar2} and \eqref{qPIchar3} from the singularity patterns, as given in \eqref{correstwo}.}\medskip

Of course, the two mappings we have chosen in these examples both only have a single open singularity pattern and one could wonder how to use Proposition \ref{prop1}  in case there are many singularity patterns. The following proposition allows us to tackle such mappings.

First we must introduce the $k$-th cyclotomic polynomial in the variable $t$, which we denote by $\phi_k(t)$:
\begin{equation}
\phi_k(t):=\prod_j (t-\iota_j),
\label{ctp}
\end{equation}
where the product runs over all primitve $k$th roots of unity, $\iota_j$.

{\prop\label{prop2} Assume that $\fie: X\to X$ is an automorphism on the space of initial conditions $X$ for a mapping with unbounded degree growth. Let $C_1, \hdots, C_m$ be (the divisor classes of) curves that appear in the open patterns for $\fie$ on $X$, let $a_1, \hdots, a_m>0$ and consider $F=\sum_{j=1}^m a_j C_j\, \in\pik{X}/\PP{X}$. Then
\begin{equation}\label{eq:notinpx}
	(\varphi^{m_0}_{*} - 1) \left( \prod^{\ell}_{j=1} \changes{\phi_{k_j}(\varphi^{m_j}_{*}) } \right) F \notin P_X
\end{equation}
for $\ell \ge 0$, $k_j \ge 2$ and $m_j \ge 1$ {$(j=0,1, \hdots, \ell)$}.
In particular, $F, (\varphi_{*} - 1) F \notin P_X$. (If $\ell=0$ the product in \eqref{eq:notinpx} is taken to be 1.)
}\medskip

{\it Proof:} If $m_0 \ge 2$, then
\begin{equation}
	t^{m_0} - 1 = (t - 1) (t^{m_0-1} + \cdots + 1)
\end{equation}
and the second factor of the right hand side decomposes into a product of cyclotomic polynomial other than $(t - 1)$.
Therefore, we may assume that $m_0 = 1$.
It follows from Lemma~\ref{lem:k1} that $(\varphi_{*} - 1) F \notin P_X$.
Thus, by Lemma~\ref{lem:k2} we have \eqref{eq:notinpx}.
Since $\varphi_{*}$ preserves $P_X$, it follows from \eqref{eq:notinpx} that $F \notin P_X$.\hfill{$\blacksquare$}\medskip

Before proving the two lemmas needed in the proof of Proposition \ref{prop2}, we first need to prove two slightly technical lemmas.

\begin{lemma}\label{lem:integrable}
Let $\fie$ correspond to an integrable mapping (case (b) in Theorem \ref{DF}).
Let $v_1 \in \operatorname{Pic} X$ be the eigenvector of $\varphi_{*}$ in the Jordan block of size 3 in Theorem \ref{DF} (b), i.e.\ $\mathbb{Q} v_1 = \ker (\varphi_{* \mathbb{Q}} - 1) \cap \operatorname{im} (\varphi_{* \mathbb{Q}} - 1)^2$, $v_1 \in \operatorname{Pic} X$ is primitive and $v_1$ is nef (see \cite{mase}).
If an irreducible curve $C \subset X$ satisfies $C^2 < 0$ and $C \cdot v_1 = 0$, then there exists $\ell > 0$ such that $\varphi^{\ell}_{*} C = C$.
\end{lemma}

{\it Proof:} 
Let $v^{\perp}_1 = \{ F \in \operatorname{Pic} X \mid F \cdot v_1 = 0 \}$ and let $L$ be the quotient lattice $v^{\perp}_1 / \mathbb{Z} v_1$.
Then from Lemma~3.6 and Lemma~3.7 in \cite{mase} it follows that the intersection form on $v^{\perp}_1$ is semi-negative definite and that its kernel is generated by $v_1$.
Hence, the bilinear form $L \times L \to \mathbb{Z}; ([u], [w]) \mapsto u \cdot w$ is well-defined and negative definite.
Since the induced action $\widetilde{\varphi}_{*} \colon L \to L$ preserves this negative definite bilinear form, there exists a positive integer $\ell$ such that $\widetilde{\varphi}^{\ell}_{*} = 1_L$ (as $\widetilde\fie_*$ is unitary and therefore acts as a (signed) permutation on the lattice).
Thus there exists $m \in \mathbb{Z}$ such that $\varphi^{\ell}_{*} C = C + m v_1$.
Suppose now that  $m \ne 0$. Then $\varphi^{\ell}_{*} C$ and $C$ are divisor classes of different irreducible curves and thus $(\varphi^{\ell}_{*} C) \cdot C \ge 0$.
If $m > 0$, then we have
\begin{equation}
	0 > (\varphi^{\ell}_{*} C)^2 = (\varphi^{\ell}_{*} C) \cdot C + m (\varphi^{\ell}_{*} C) \cdot v_1.
\end{equation}
However, the right hand side is nonnegative since $v_1$ is nef, which leads to a contradiction.
On the other hand if $m < 0$ we have
\begin{equation}
	0 \le (\varphi^{\ell}_{*} C) \cdot C - m (\varphi^{\ell}_{*} C) \cdot v_1 = C^2 < 0,
\end{equation}
which is also a contradiction.
Hence, we have $m = 0$ and $\varphi^{\ell}_{*} C = C$.\hfill{$\blacksquare$}

\begin{lemma}\label{lem:nonintegrable}
Let $\fie$ correspond to a nonintegrable mapping (case (c) in Theorem \ref{DF}).
Let $v \in \operatorname{Pic}_{\mathbb{R}} X$ be the dominant eigenvector of $\varphi_{*}$, i.e.\ $v$ corresponds to the eigenvalue that gives the dynamical degree and is nef (see\cite{mase}).
If $F \in \operatorname{Pic} X$ satisfies $F \cdot v = 0$, then there exists $\ell > 0$ such that $\varphi^{\ell}_{*} F = F$.
\end{lemma}

{\it Proof:} 
Let $L = \{ G \in \operatorname{Pic} X \mid G \cdot v = 0 \}$.
From Lemma~4.15 in \cite{mase} it follows that the intersection form is negative definite on $L$.
As in the proof of Lemma \ref{lem:integrable}, since $\varphi_{*}$ preserves $L$ there exists a positive integer $\ell$ such that $(\varphi_{*}|_L)^{\ell} = 1_L$, which implies $\varphi^{\ell}_{*} F = F$.\hfill{$\blacksquare$}\medskip

Using \changes{these} two technical lemmas, we can now prove the remaining ingredients needed in the proof of Proposition \ref{prop2}.

\begin{lemma}\label{lem:k1}
Let $C_1, \ldots, C_m \subset X$ appear in open chains and let $a_1, \ldots, a_m > 0$.
Let $F = \sum_j a_j C_j$.
Then
\begin{equation}
	(\varphi_{*} - 1) F \notin P_X.
\end{equation}
\end{lemma}

{\it Proof:} 
We consider first the nonintegrable case.
Since $\mu_{\overline{\varphi}_{*}}(t)$ does not have a factor $(t - 1)$ (see Corollary \ref{corol2}) it is sufficient to show that $F \notin P_X$.
Assume that there exists $\ell > 0$ such that $\varphi^{\ell}_{*} F = F$.
Let $v \in \operatorname{Pic}_{\mathbb{R}} X$ be the dominant eigenvector of $\varphi_{*}$.
Using
\begin{equation}
	v \cdot F = (\varphi^{\ell}_{*} v) \cdot (\varphi^{\ell}_{*} F)
	= \lambda^\ell v \cdot F,
\end{equation}
we have $v \cdot F = 0$ since $\lambda>1$.
Furthermore, as $v$ is nef, $a_j > 0$ and
\begin{equation}
	\sum a_j v \cdot C_j = 0,
\end{equation}
$v \cdot C_j$ must be zero for all $j$.
However, from Lemma~\ref{lem:nonintegrable} we must then have that $C_j \in P_X$, which is a contradiction since $C_j$ appears in an open chain.
Therefore, we have $F \notin P_X$ in the nonintegrable case.

Next, we consider the integrable case.
Assume that $(\varphi_{*} - 1) F \in P_X$.
Then, there exist $G \in \operatorname{Pic} X$ and $\ell > 0$ such that
\begin{equation}
	(\varphi_{*} - 1) F = G, \quad
	\varphi^{\ell}_{*} G = G.
\end{equation}
Take $v_2, v_3 \in \operatorname{Pic}_{\mathbb{Q}} X$ such that
\begin{equation}
	\varphi_{*} v_3 = v_3 + v_2, \quad
	\varphi_{*} v_2 = v_2 + v_1.
\end{equation}
Using
\begin{equation}
	v_2 \cdot G = (\varphi^{\ell}_{*} v_2) \cdot (\varphi^{\ell}_{*} G)
	= (v_2 + \ell v_1) \cdot G
	= v_2 \cdot G + \ell v_1 \cdot G,
\end{equation}
we have $G \cdot v_1 = 0$.
Since
\begin{equation}
	v_3 \cdot G = (\varphi^{\ell}_{*} v_3) \cdot (\varphi^{\ell}_{*} G)
	= \left( v_3 + \ell v_2 + \frac{\ell(\ell - 1)}{2} v_1 \right) \cdot G
	= v_3 \cdot G + \ell v_2 \cdot G,
\end{equation}
we  also have $G \cdot v_2 = 0$.
Then, using
\begin{equation}
	v_2 \cdot F = (\varphi_{*} v_2) \cdot (\varphi_{*} F)
	= (v_2 + v_1) \cdot (F + G)
	= v_2 \cdot F + v_1 \cdot F,
\end{equation}
we find that $v_1 \cdot F = 0$.
Since $v_1$ is nef, $a_j > 0$ and
\begin{equation}
	\sum a_j v_1 \cdot C_j = 0,
\end{equation}
$v_1 \cdot C_j$ must be zero for all $j$.
However, it then follows from Lemma~\ref{lem:integrable} that $C_j \in P_X$, which is a contradiction and we find that $(\varphi_{*} - 1) F \notin P_X$ also in this case. \hfill{$\blacksquare$}

\begin{lemma}\label{lem:k2}
If $F \in \operatorname{Pic} X \setminus P_X$, then
\begin{equation}
	\prod^{\ell}_{j=1} \phi_{k_j}(\varphi^{m_j}_{*}) F \notin P_X
\end{equation}
for $\ell \ge 0$, $k_j \ge 2$ and $m_j \ge 1$ $(j=1, \hdots, \ell)$.
\end{lemma}

{\it Proof:} 
Since each $\phi_{k_j}(t^{m_j})$ decomposes into a product of cyclotomic polynomials other than $(t - 1)$, we may assume that $m_1 = \cdots = m_{\ell} = 1$.
For $\ell = 1$, it follows from the classification given in Corollary \ref{corol2} that $\mu_{\overline{\varphi}_{*}}(t)$ is not divisible by $\phi_{k_1}(t)$ as $k_1 \ge 2$.
Therefore, $\phi_{k_1}(\overline{\varphi}_{*})$ cannot annihilate any nonzero element in $\pik{X}/\PP{X}$.
Hence, we have $\phi_{k_1}(\varphi_{*}) F \notin P_X$.
Since
\begin{equation}
	\left( \prod^{\ell}_{j=1} \changes{\phi_{k_j}(\varphi_{*}) }\right) F
	= \changes{\phi_{k_{\ell}}(\varphi_{*})  } \left( \prod^{\ell-1}_{j=1} \changes{\phi_{k_j}(\varphi_{*}) } \right) F,
\end{equation}
the proof is done by induction on $\ell$ in case $\ell \ge 2$.\hfill{$\blacksquare$}\medskip

We shall \changes{explain} the use of Proposition \ref{prop2} on some illustrative examples in the next section.

{\remark Proposition 2 and related Lemmas also hold for nonautonomous confining mappings by replacing $\fie_*$ by a Cremona isometry of infinite order on $\pik{X}$. All proofs can then be used as such.}

\section{Singularity patterns and integrability}\label{examps}

We start by giving an example of a family of confining mappings with only one open singularity pattern, but for which Proposition \ref{prop2} is nevertheless crucial in deriving the value of their dynamical degrees.

{\example~\!\!} Let us consider the family of confining mappings \cite{kankiHVext} given by the equation
\begin{equation}
x_{n+1} + x_{n-1} = x_n + \frac{1}{x_n^k},
\label{map2k}
\end{equation}
where $k$ is a positive even integer. At $k=2$ this family comprises the archetypical example of a nonintegrable confining mapping, the Hietarinta-Viallet mapping \cite{HV}. 

The mappings in the family \eqref{map2k} all have the same two singularities: at $x_n=0$ and $\infty$. The first singularity is confined, with an open singularity pattern
\begin{equation}
(0~~\infty^k~~\infty^k~~0),
\label{map2ksipa}
\end{equation}
whereas the second singularity corresponds to the cyclic singularity pattern 
\begin{equation}
(x_0~\infty~\infty).
\label{map2kcycl}
\end{equation}
Each mapping in this family, re-interpreted as a mapping on $\cpp$ of the form \eqref{mapCFcpp}, can therefore be regularized to an automorphism $\fie: X\to X$ on a rational surface $X$. On this surface we then have the open pattern of curves
\begin{equation}
\{y=0\} \to C_1 \to C_2 \to C_3 \to \{x=0\},
\end{equation}
where $\{y=0\}$ ($\{x=0\}$) are the strict transforms of $[y=0]$ ($[x=0]$) on $\cpp$, and where \changes{the curves $C_1, C_2$ and $C_3$, respectively, correspond to $\footnotesize\ve{0}{\infty^k}, \ve{\infty^k}{\infty^k}$ and $\footnotesize\ve{\infty^k}{0}$ (in the sense of Remark \ref{zenotation}).}
From Lemma \ref{trivlem} we have that \changes{classes}  $C_1, C_2, C_3\neq 0 \mod\PP{X}$.

Just as we did in Examples \ref{ex41} and \ref{ex42}, we shall use linear equivalence on $\pik{X}$ to express \changes{$H_y$, the class of the total transform of $[y=0]\subset\cpp$,} in various forms:
\begin{align}
H_y =~& \{y=0\} + C_3 \mod\PP{X} ~= \big(\fie_*^{-1} + \fie_*^2\big)\,C_1 \mod\PP{X}\\
=~& k (C_1 + C_2) \mod\PP{X} ~= k \big(\fie_* +1\big)\, C_1 \mod\PP{X},
\end{align}
where the curves (and their multiplicities in the blow-up structure)  involved in these expressions can be read off directly from the singularity pattern  \eqref{map2ksipa}.
From these expressions we obtain
\begin{equation}
\big( \fie_*^3 - k \fie_*^2 - k \fie_* + 1 \big)\, C_1 = 0 \mod\PP{X},\label{tobereduced}
\end{equation}
and, as the polynomial $(t^3 - k t^2 - k t + 1) \equiv (t+1) \left(t^2 -(k+1) t +1\right)$ does not have a root equal to 1, we conclude from Proposition \ref{prop1} that all the mappings in this family must be nonintegrable.

However, Proposition \ref{prop2} allows us to reach an even stronger conclusion. 
Since $(t+1)$ is the cyclotomic polynomial $\phi_2(t)$ (cf. formula \eqref{ctp}), we know that $F:=\big(\fie_*+1\big)\,C_1\neq0\mod\PP{X}$. From relation \eqref{tobereduced} we thus have 
\begin{equation}
\big( \fie_*^2 - (k+1) \fie_* + 1 \big)\, F = 0 \mod\PP{X},
\end{equation}
for $F \notin P_X$, and Proposition \ref{prop1} then tells us that $\psi(t)=\left(t^2 -(k+1) t +1\right)$, which is irreducible for $k\geq 2$, must coincide with the minimal polynomial $\mu_{\pfie}(t)$. 
The largest root of $\left(t^2 -(k+1) t +1\right)$,
\begin{equation}
\lambda_* = \frac{k+1 + \sqrt{(k+1)^2-4\,}}{2},
\end{equation}
is therefore also the largest eigenvalue of $\fie_*$ and yields the value of the dynamical degree for each member of the family \eqref{map2k} \cite{kankiHVext} (see also \cite{redemption}). Note that for $k=2$, this indeed coincides with the known value of $(3+\sqrt{5})/2$ for the dynamical degree of the Hietarinta-Viallet mapping \cite{HV,takenawaHV}.\medskip

Next we give an example of a mapping with a longer open singularity pattern.

{\example~\!\!} Consider the birational mapping on $\cpp$ of the form \eqref{mapCFcpp}, defined by the equation
\begin{equation}
x_{n+1} x_{n-1} = \frac{x_n-b}{x_n-1}\qquad (b\neq 0,1).
\label{ex52}
\end{equation}
This mapping has two singularities, at $b$ and $1$. The singularity at $x_n=b$ is confined, with open singularity pattern
\begin{equation}
(b~~0~~1~~\infty^2~~1~~0~~b),
\label{qPIISsipa}
\end{equation}
and that at $x_n=1$ is part of a cyclic singularity pattern
\begin{equation}
(x_0~1~\infty~1~x_0'),
\label{qPIIScycl}
\end{equation}
where $x_0'=x_0/(x_0-1)$ 
. As the mapping is confining it has a space of initial conditions, i.e. it can be lifted to an automorphism $\fie$ on a rational surface $X$, on which we have the open pattern
\begin{equation}
\{y=b\} \to C_1 \to C_2 \to C_3 \to C_4 \to C_5 \to C_6 \to \{x=b\},
\end{equation}
in which curly brackets denote the strict transforms of the indicated curve. Since the curves indicated as $C_j$ appear in an open pattern, we have from Lemma \ref{trivlem} that $C_j\neq0\mod\PP{X}~(j=1,\hdots,6)$. From linear equivalence on $\pik{X}$ we find for the \changes{(class of the)} total transform of the curve $[y=b]$ on $\cpp$:
\begin{align}
H_y =~& \{y=b\} + C_6 \mod\PP{X} ~= \big(\fie_*^{-1} + \fie_*^5\big)\, C_1 \mod\PP{X}\label{ex52eq1}\\
=~& C_1 + C_5 \mod\PP{X} ~= \big(\fie_*^4 + 1\big)\, C_1 \mod\PP{X}\label{ex52eq2}\\
=~& C_2 + C_4 \mod\PP{X} ~= \big(\fie_*^3 + \fie_*\big)\, C_1 \mod\PP{X}\label{ex52eq3}\\
=~& 2 C_3 \mod\PP{X} ~= 2 \fie_*^2\, C_1 \mod\PP{X},\label{ex52eq4}
\end{align}
where, besides the open singularity pattern \eqref{qPIISsipa}, we have used the fact that $ \{y=0\},  \{y=1\}$ and $ \{y=\infty\}$ lie in $\PP{X}$ (Lemma \ref{trivlem}).

{\remark Instead of invoking Lemma \ref{trivlem}, one could of course also check explicitly whether \changes{the classes of} these strict transforms lie in $\PP{X}$. For $\{y=1\}$ this is immediate as it is part of a cyclic pattern on $X$ induced by the cyclic singularity pattern \eqref{qPIIScycl}. On the other hand, the strict transforms $\{y=0\}$ and $\{y=\infty\}$ are part of a cyclic pattern on $X$ that does not correspond to a singularity pattern:
\begin{equation}
( x_0~\infty~x_0'~0 ),
\end{equation}
where $x_0'=1/x_0$.}\medskip

From the above relations (\ref{ex52eq1}$\sim$\ref{ex52eq4}) we can derive 6 different relations. However, as we shall see, it suffices to just choose one. For example, using \eqref{ex52eq3} and \eqref{ex52eq4}, we obtain
\begin{equation}
\big( \fie_*^2 - 2 \fie_* +1\big)\,C_1 \equiv \big( \fie_*-1\big)^2\, C_1 = 0\mod\PP{X}.
\end{equation}
The ensuing polynomial $(t-1)^2$ does not have any roots different from 1 and Proposition \ref{prop1} therefore tells us that the mapping is integrable: its dynamical degree is 1. Note that the same conclusion could have been reached from any other of the relations we could have build.

Next we give an example with more than one open singularity pattern.
{\example~\!\!} One might be tempted to think of mapping \eqref{ex52} as a mere special case of the equation
\begin{equation}
x_{n+1} x_{n-1} = a \frac{x_n-b}{x_n-1}\qquad (a,b\neq 0,1,~ a\neq b),
\label{ex53}
\end{equation}
but geometrically speaking these two mappings are quite different, as is clear immediately from their singularity analysis.

Just as mapping \eqref{ex52},  mapping \eqref{ex53} also has two singularities, at $b$ and $1$, but both are now confined with open singularity patterns \cite{rodone}:
\begin{gather}
(1~~\infty~~a~~0~~b)\\
(b~~0~~a~~\infty~~1),
\end{gather}
but there are no cyclic singularity patterns. The mapping \eqref{ex53}, being confining, therefore has a space of initial conditions, $X$, in which the above two singularity patterns induce open patterns of the form:
\begin{gather}
\{y=1\} \to C_1 \to C_2 \to C_3 \to C_4 \to \{x=b\}\\
\{y=b\} \to C_1' \to C_2' \to C_3' \to C_4' \to \{x=1\}.
\end{gather}
Using linear equivalence on $\pik{X}$ to express $H_y$ \changes{(the class of the total transform of $[y=0]\subset\cpp$)} in various forms, we obtain:
\begin{align}
H_y = \{y=1\} + C_4' = \{y=b\} + C_4 = C_1 + C_3' = C_3 + C_1' = C_2 + C_2' \mod\PP{X},
\end{align}
the \changes{classes of the} strict transforms $\{y=\infty\}, \{y=0\}$ and $\{y=a\}$ lying in $\PP{X}$ due to Lemma \ref{trivlem} (cf. \cite{rodone} for the cyclic patterns these curves belong to). These expressions can be re-expressed in terms of $\fie_*: \pik{X}\to\pik{X}$ as
\begin{align}
H_y = \fie_*^3\,C_1' + \fie_*^{-1}\, C_1 =\fie_*^3\,C_1 + \fie_*^{-1}\,C_1' = \fie_*^2\,C_1' + C_1 = \fie_*^2\,C_1 + C_1' = \fie_* (C_1 + C_1') \mod\PP{X},
\end{align}
from which we can, for example, derive:
\begin{gather}
\big(\fie_*^2+1\big) (C_1 + C_1') = 2 \fie_* (C_1 + C_1') \mod\PP{X}\label{useful1}\\
\Leftrightarrow\qquad \big(\fie_* - 1\big)^2 (C_1 + C_1') = 0 \mod\PP{X}.
\end{gather}
From Proposition \ref{prop2} we know that $C_1 + C_1'\neq 0 \mod\PP{X}$ and hence, because the polynomial (in $\fie_*$) that annihilates $C_1 + C_1$ in $\pik{X}/\PP{X}$ does not possess any roots other than 1, we find from Proposition \ref{prop1} that this mapping is integrable as well. \medskip

We end this list of examples with a rather extreme case of a mapping with several singularity patterns.

{\example~\!\!} Let us consider the confining mapping $f: \cpp\dashrightarrow\cpp$, defined by the equation \cite{mimura},
\begin{equation}
x_{n+1} x_{n-1} = \frac{x_n^4-1}{x_n^4+1},
\label{ex54}
\end{equation}
which has 8 open singularity patterns:
\begin{gather}
(\pm1~~0~~\mp1),\qquad (\pm i~~0~~\pm i), \qquad (\pm r~~\infty~~\mp i r),\qquad (\pm i r~~ \infty~~\mp r),
\end{gather}
where $r$ is the square root of $i$, i.e. $r=\exp i\pi/4$. In this case we have 16 special curves $A_1^+, \cdots, D_2^-$ on the rational surface $X$ on which the mapping is regularized, that appear in the open patterns
\begin{gather}
\{y=\pm1\} \to A_1^\pm \to A_2^\pm \to \{x=\mp1\}\\
\{y=\pm i\} \to B_1^\pm \to B_2^\pm \to \{x=\pm i\}\\
\{y=\pm r\} \to C_1^\pm \to C_2^\pm \to \{x=\mp i r\}\\
\{y=\pm ir\} \to D_1^\pm \to D_2^\pm \to \{x=\mp r\}.
\end{gather}
The \changes{classes of the} strict transforms $\{y=0\}$ and $\{y=\infty\}$ are part of $\PP{X}$ (because of Lemma \ref{trivlem}) and we therefore find for the \changes{class $H_y$ of the} total transform of $[y=0]\subset\cpp$ that \changes{(modulo $\PP{X}$)}
\begin{align}
H_y =~& \{y=1\} + A_2^- = \{y=-1\} + A_2^+ = \{y=i\} + B_2^+ = \{y=-i\} + B_2^-\\
       =~& \{y=r\} + D_2^- = \{y=-r\} + D_2^+ = \{y=i r\} + C_2^- = \{y=-i r\} + C_2^+\\
       =~& A_1^+ + A_1^- + B_1^+ + B_1^- = C_1^+ + C_1^- + D_1^+ + D_1^- ,
\end{align}
from which we derive
\begin{align}
&\big(\fie_* + \fie_*^{-1}\big)\, F = 4 F\mod\PP{X}\label{useful2}\\
\Leftrightarrow\qquad &\big(\fie_*^2 - 4 \fie_* + 1\big)\, F = 0\mod\PP{X},
\end{align}
where $F := A_1^+ + A_1^- + B_1^+ + B_1^- + C_1^+ + C_1^- + D_1^+ + D_1^- \neq 0 \mod\PP{X}$ (Proposition \ref{prop2}).

As $(t^2-4t+1)$ does not contain a factor $(t-1)$, we conclude from Proposition \ref{prop1} that this mapping is nonintegrable. Moreover, since this polynomial 
is irreducible over $\mathbb{Z}$ (and therefore over $\mathbb{Q}$), it follows also that the minimal polynomial for $\fie_*:\pik{X}\to\pik{X}$, $\mu_{\fie_*}(t)$, has to contain it as a factor and hence, that the largest eigenvalue of $\fie_*$ must be equal to the largest root of $(t^2-4t+1)$, i.e.: $2+\sqrt{3}$. This is then the value of the dynamical degree for the mapping \eqref{ex54} (see also \cite{redemption}).

\section{Conclusion}\label{conc}
In this paper we have shown how, for confining three point mappings with unbounded degree growth, the algebro-geometric properties of their space of initial conditions can be used to decide on their integrability in a very simple way. As explained, the method we propose here is in fact an algebro-geometric reformulation of the express method we introduced in \cite{rodone} for such mappings.

Although the method relies on geometric notions and on deep properties of the rational surfaces on which the mappings are regularized, it does not require any detailed calculation of actual blow-ups (or blow-downs), nor does its application require any genuine algebro-geometric  expertise on the side of its practitioner. It is sufficient to perform a standard, but careful, singularity analysis of the mapping, taking into account the multiplicities in each singularity pattern, followed by a straightforward application of Propositions \ref{prop1} and \ref{prop2} of section \ref{mres}.
If the result of this analysis is that a given mapping is integrable, its dynamical degree is of course known to be 1, by definition. On the other hand, for nonintegrable mappings, in all cases known to us, the method we presented here also yields the exact value of the dynamical degree.

There are however a few theoretical and practical limitations to the method, as it stands. A first, obvious, limitation is that the method can only be applied to birational mappings on $\cpp$. This is, unfortunately, a real limitation: as long as there is no equivalent of the structure theorem for birational mappings on $\cpp$ (Theorem \ref{DF}), for mappings on $\mathbb{P}^n(\mathbb{C})$, there is no hope of developing a similar framework for establishing the integrability of higher order mappings. However, the fact that we restricted our discussion to autonomous mappings is not essential: our approach works perfectly for confining nonautonomous mappings as well. 

A second limitation is one we imposed for practical reasons: throughout the paper we restricted the discussion to three point mappings, for the simple reason that for such mappings the singularity patterns can all be expressed in a single coordinate on $\cpp$ (in our case the $y$-coordinate), which greatly simplifies the singularity analysis. We do not claim that it is impossible to generalize our method to more general birational mappings on $\cpp$, just that the practical implementation  of such a generalized method might be hopelessly complicated. As the singularities of a more general mapping can arise anywhere in the projective plane, computing useful relations between curves in the space of initial conditions for such mappings, as e.g. in equations \eqref{useful1} or \eqref{useful2}, could become prohibitively difficult.

A third problem is that there is actually no guarantee that the singularity patterns, as obtained from the mapping in its original form, will actually yield any meaningful relations at all. This is a possibility we already pointed out when we introduced the express method. In \cite{rodone} we give an example of a discrete  Painlev\'e equation with E$_8^{(1)}$ symmetry, that has 8 short open singularity patterns. In fact, these patterns are too short to yield any useful relations that would allows us to implement the express method (or, for that matter, the present method). As we explain in that paper, however, it is possible to replace the original mapping by a coupled one (which implies the original mapping in one of the dependent variables), thereby effectively lengthening the singularity patterns so that they do become useful.
 
 Another problem concerns Proposition \ref{prop1}, or more precisely the criterion used in it to distinguish between integrable and nonintegrable cases, which is not a strict dichotomy. Although we have never encountered such a mapping, there might exist examples for which the sole polynomial $\psi(t)$ (of Proposition \ref{prop1}) that arises in our analysis, contains a factor $(t-1)^2$ {\it as well as} an irreducible factor $g(t)$ (with $\deg g(t)\geq2$) with a root greater than 1. In which case Proposition \ref{prop1} (even in combination with Proposition \ref{prop2}) does not allow us to conclude anything. We do not know of a single example of such a mapping, but for now we see no reason why such a mapping could not exist.
 
 A last restriction is that the method, obviously, only applies to confining mappings. However, in \cite{rodnon} (cf. also Appendix 2), we have presented a version of the express method adapted to non-confining mappings. It is known \cite{dillerfavre} that in such a case the mapping cannot be lifted to an automorphism on a rational surface with a finite number of blow-ups (or blow-downs), and that if such a mapping is not linearisable, it must be nonintegrable. The method we describe in \cite{rodnon} allows us to treat both scenarios but, for the time being, we have not succeeded in reformulating it in a suitable algebro-geometric framework. We hope to have some success in this direction in the near future.

\section*{Acknowledgements}
RW and TM would like to acknowledge support from the Japan Society for the Promotion of Science (JSPS),  through JSPS grants number 18K03355 and 18K13438, respectively.


\section*{Appendix 1: Halburd's method for calculating degrees}\label{app1}
In section \ref{secqPI}, the information contained in the cyclic pattern \eqref{sipacyclqPI} was completely negelected in  our analysis of the degree growth of mapping \eqref{qPI}. As we saw in section \ref{mres}, this is justified if one only wants to assess the integrability (or non-integrability) of the mapping. However, if as in Halburd's method \cite{rodzero} the aim is to calculate the exact degree of each iterate of the mapping -- in our case starting from a generic value for $x_0$ and from $x_1=z\in\cp$ -- as a rational function in $z$, then the information contained in cyclic patterns such as \eqref{sipacyclqPI} is crucial.

For example, in the case of mapping \eqref{qPI} it is easy to see that the degree $d_n$ of the nth iterate $f_n(z)$ of the mapping,
\begin{equation}
d_n:= \deg_z f_n(z),
\end{equation}
calculated as the number of pre-images $d_n(-1)$ of the value -1, 
\begin{equation}
d_n(-1):=\#\{z\in\cp\big \vert ~ f_n(z)=-1\},
\end{equation}
counted with multiplicities, is given by
\begin{equation}
d_n = d_n(-1) = M_{n-4} + M_n.
\label{app1rel1}
\end{equation}
Here, as in section \ref{secqPI}, $M_n$ denotes the number of (spontaneous) occurrences of the value -1 at the nth itearte of the mapping. The values $0$ and $\infty$ however appear in both the open pattern \eqref{sipaqPI} as well as in the cyclic one \eqref{sipacyclqPI}. More precisely, a value 0 appears 1 step as well as 3 steps after a value -1 in the open pattern, but also spontaneously every first or third step (as is clear from the substring $ \cdots ~x_0~ 0~ \infty^2~0~\cdots $ in the cyclic pattern  \eqref{sipacyclqPI}) as well as every second step in the cyclic pattern (as is clear from the substring $ \cdots ~x_0'~ \infty~0~\cdots $ in  \eqref{sipacyclqPI}). Hence, the degree of $f_n(z)$, calculated as the number of pre-images of 0, is given by
\begin{equation}
d_n = d_n(0) = M_{n-1} + M_{n-3} + \dfrac{3- (-1)^n - i^n - (-i)^n}{4},
\label{app1rel2}
\end{equation}
where $i=\sqrt{-1}$. Similarly, it can be seen that the degree of $f_n(z)$ calculated as the number of pre-images of $\infty$ is given by
\begin{equation}
d_n = d_n(\infty) = 2 M_{n-2} + 1- \dfrac{ i^n + (-i)^n}{2},
\label{app1rel3}
\end{equation}
as an $\infty$ appears with multiplicity 2 two steps after a -1 in the open pattern, and spontaneously, with multiplicity 1 at every first and third step, as well as with multiplicity 2 at every second step in the cyclic pattern.

Solving the simplest equation among the three possible ones, $d_n(0)=d_n(\infty)$,
\begin{equation}
M_{n+1} - 2 M_n + M_{n-1} = \frac{1+(-1)^n + i^n + (-1)^n}{4},
\end{equation}
for $n\geq0$, with $M_0=0, M_1=1$, and then using any of the relations \eqref{app1rel1}, \eqref{app1rel2} or \eqref{app1rel3} to calculate the degree, we obtain
\begin{equation}
d_n = \frac{n^2}{4} + \frac{5}{8} - \frac{(-1)^n}{8} - \frac{i^n+(-i)^n}{4}\qquad (n\geq 0)
\end{equation}
for the degree in $z$ of the nth iterate $f_n(z)$ of mapping \eqref{qPI}. This degree is quadratic in $n$ and hence, the dynamical degree $\lim_{n\to+\infty} d_n^{1/n} = 1$, as it should be for a QRT mapping.

Although slightly involved, the above calculation is quite straightforward once the singularity patterns have been established.
The case of mapping \eqref{mapk} however should serve as a cautionary note. As this mapping only has a single singularity, it obviously also only possesses one singularity pattern, an open one. Nonetheless does exist a cycle
\begin{equation}
x_0~ \infty~ x_0' ~ \infty ~\cdots
\label{mapkcycle}
\end{equation}
(where $x_0'=-x_0$) which clearly amounts to a simple exchange of curves in $\cpp$. I.e.: this cycle does not contain any singularities and is therefore extraneous to the singularity analysis of the mapping \eqref{mapk}. Note that, because it does not contain any singularities, such a cycle might be quite hard to discover for more complicated mappings. Nevertheless, the information contained in this cycle is crucial if one wants to calculate the exact degree sequence for the iterates of the mapping using Halburd's method. 

Obviously, the degree $d_n$ of $f_n(z)$ is given as the number of pre-images of the value $0$ by
\begin{equation}
d_n=d_n(0)=Z_{n-2} + Z_n,
\end{equation}
as explained in section \ref{secmapk}. On the other hand, for the number of pre-images of $\infty$ one finds
\begin{equation}
d_n= d_n(\infty) = k Z_{n-1} + \frac{1-(-1)^n}{2},
\end{equation}
since an $\infty$ can also occur at every odd step in the cycle \eqref{mapkcycle}. Hence we find the equation
\begin{equation}
Z_n+ Z_{n-2} = k Z_{n-1} + \frac{1-(-1)^n}{2},
\end{equation}
for $k\in2\mathbb{Z}_{\geq1}$ and $n\geq0$, with initial conditions $Z_0=0, Z_1=1$. The solution can be expressed as
\begin{equation}
Z_n=\left\{
\begin{matrix}
\dfrac{1}{2(2-k)} - \dfrac{(-1)^n}{2(2+k)} + \dfrac{\lambda_*^{n+1}+\lambda_*^{-n-1}}{k^2-4} & (k>2)\\[-2mm]\\
\dfrac{n(n+2)}{4} + \dfrac{1-(-1)^n}{8} & (k=2)
\end{matrix}
\right.
\end{equation}
where $\lambda_*=(k+\sqrt{k^2-4\,})/2$, as in \eqref{dydegk}. This yields the following expressions for the degree $d_n$ of $f_n(z)$:
\begin{equation}
d_n=\left\{
\begin{matrix}
\dfrac{1}{(2-k)} - \dfrac{(-1)^n}{(2+k)} + \dfrac{k (\lambda_*^{n}+\lambda_*^{-n})}{k^2-4}& (k>2)\\[-2mm]\\
\dfrac{n^2}{2} + \dfrac{1-(-1)^n}{4} & (k=2)
\end{matrix}
\right.
\end{equation}
which shows that the dynamical degree $\lim_{n\to\infty} d_n^{1/n}$ is indeed 1 when $k=2$, and $\lambda_*$ when $k\in2\mathbb{Z}_{\geq2}$.

\section*{Appendix 2: the express method in a non-confining case}\label{app2}
When $k$ is an odd positive integer, the mapping \eqref{mapk} does not possess the singularity confinement property. As a matter of fact, as the sign of the nonlinearity is not fixed in this case, no cancellations occur and the degree growth is maximal when $k>1$.

On the other hand, when $k=1$, \eqref{mapk} has linear degree growth and is in fact linearisable. This is easily seen by calculating the update $n\to n+1$ of the quantity $x_{n-1} x_n -1/2$:
\begin{equation}
x_n x_{n+1} -\frac{1}{2} = x_n \Big(-x_{n-1} + \frac{1}{x_n}\Big) -\frac{1}{2} \equiv - \Big(x_{n-1} x_n -\frac{1}{2}\Big),
\end{equation} 
from which the linearisability is obvious.

Although for $k>1$, odd, the singularity that arises for \eqref{mapk} at $x_n=0$ is not confined, it is possible to analyse it in detail. For example, the singularity corresponds to the infinitely repeating singularity pattern
\begin{equation}
\sipa{0^1~ \infty^k~0^1~ \infty^k~0^1~ \infty^k~0^1~ \infty^k~0^1~ \infty^k~\cdots },
\end{equation}
from which it is actually possible, if we also take into account the cyclic pattern \eqref{mapkcycle}, to obtain the exact degree for the iterates of \eqref{mapk}  in this case as well. Following the procedure outlined in \cite{rodnon}, we express the degree of the nth iterate $f_n(z)$ (obtained from a generic $x_0$ and $x_1=z$) in terms of the number of spontaneous occurrences of the value 0 (denoted $Z_n$) as:
\begin{equation}
d_n=d_n(0) = Z_n + Z_{n-2} + Z_{n-4} + Z_{n-6} + \cdots,
\end{equation}
where we take $Z_{n\leq0}=0$.

This has to match the degree calculated as the number of pre-images of $\infty$, $d_n(\infty)$, which is given by
\begin{equation}
d_n=d_n(\infty) = \frac{1-(-1)^n}{2} + k (Z_{n-1} + Z_{n-2} + Z_{n-5} + \cdots),
\end{equation} 
because of the contribution of the cycle \eqref{mapkcycle}. We thus obtain from $d_n(0)=d_n(\infty)$,
\begin{equation}
\sum_{k=0}^{+\infty} Z_{n-2k} = \frac{1-(-1)^n}{2} + k \sum_{k=0}^{+\infty} Z_{n-2k-1},
\label{app2eq}
\end{equation}
the  characteristic equation for the homogeneous part of which can be explicitly calculated if we assume there exists a caracteristic root $\lambda>1$:
\begin{equation}
\sum_{k=0}^{+\infty} \lambda^{n-2k} = k \sum_{k=0}^{+\infty} \lambda^{n-2k-1}\qquad\Leftrightarrow\qquad \frac{\lambda^2}{\lambda^2-1} = \frac{k}{\lambda} \frac{\lambda^2}{\lambda^2-1}.
\end{equation}
This obviously yields $\lambda=k$, which is indeed greater than 1 if $k>1$.

The full equation \eqref{app2eq} can be solved (for $k\in\mathbb{Z}_{\geq3}$, odd) as:
\begin{equation}
d_n=\left\{
\begin{matrix}
k \dfrac{k^{2\ell}-1}{k^2-1}& (n=2\ell)\\[-2mm]\\
\dfrac{k^{2\ell}-1}{k^2-1} & (n=2\ell-1)
\end{matrix}
\qquad (\ell\geq1)
\right.
\label{app2sol}
\end{equation}
which clearly shows that the dynamical degree of the mapping in this case is indeed $k$. All these mappings therefore have exponential degree growth and are nonintegrable. Note that the solution \eqref{app2sol} still has meaning at the limit $k\to 1$, yielding
\begin{equation}
d_{2\ell-1}=d_{2\ell} =\ell\qquad (\ell\geq1),
\end{equation}
which is indeed the correct value for $d_n=\deg_z f_n(z)$ in the linearisable case $k=1$.

\end{document}